\documentclass [preprint, 11pt]{elsart}
\usepackage{amsmath}
\usepackage{amssymb}
\usepackage{amssymb}
\usepackage{graphicx}
\usepackage{amsopn}
\usepackage{amsmath}
\usepackage{amsopn}
\usepackage{amsfonts}
\usepackage{amsbsy}

\newtheorem{proposition}{Proposition}

\newtheorem{theorem}{Theorem}
\newtheorem{definition}{Definition}

\input{tcilatex}

\begin{document}

\begin{frontmatter}

% Title, authors and addresses

% use the thanksref command within \title, \author or \address for footnotes;
% use the corauthref command within \author for corresponding author footnotes;
% use the ead command for the email address,
% and the form \ead[url] for the home page:
%\title{Time distance based computation of the state space of preemptive real time
%systems. }
% \thanks[label1]{}
% \author{Name\corauthref{cor1}\thanksref{label2}}
%

% \thanks[label2]{}
% \corauth[cor1]{}
% \address{Address\thanksref{label3}}
% \thanks[label3]{}

\title{Time distance based computation of the state space of preemptive real time
systems. }

% use optional labels to link authors explicitly to addresses:
% \author[label1,label2]{}
% \address[label1]{}
% \address[label2]{}

\author{Abdelkrim .Abdelli}
\address{LSI laboratory- Computer Science Faculty- USTHB university \\
          BP 32 El alia Bab-ezzouar Algiers Algeria.} \ead{Abdelli@lsi-usthb.dz}
          \ead[url]{http://www.lsi-usthb.dz/index.php?page=page32}
\begin{abstract}
We explore in this paper a novel approach that builds an
overapproximation of the state space of preemptive real time
systems. Our graph construction extends the expression of a class to
the time distance system that encodes the quantitative properties of
past fired subsequences. This makes it possible to restore relevant
time information that is used to tighten still more the DBM
overapproximation of reachable classes. We succeed thereby to build
efficiently tighter approximated graphs which are more appropriate
to restore the quantitative properties of the model. The simulation
results show that the computed graphs are of the same size as the
exact graphs while improving by far the times needed for their
computation.
\end{abstract}

\end{frontmatter}

% main text
\hfill \hfill

\section{Introduction \qquad}

Nowadays, real-time systems are becoming more and more complex and are often
critical. Generally, these systems consist of several tasks that are timely
dependent, interacting and sharing one or more resources (e.g processors,
memory). Consequently, the correctness proofs of such systems are demanding
much theory regarding their increasing complexity. We need, for instance, to
consider formal models requiring the specification of time preemption;
concept where execution of a task may be stopped for a while and later
resumed at the same point. This notion of suspension implies to extend the
semantics of timed clocks in order to handle such behaviors. For this
effect, the concept of \textit{stopwatch} has been introduced while many
models have been defined, as for instance, hybrid automata ($LHA$) \cite%
{Alur}, stopwatch automata ($SWA$) \cite{Cassez}, Network of Stopwatch Automta (NSA) \cite{Glon2018}, and timed automata with priorities \cite{Pimk2021}. Time Petri nets ($TPN$)
have also been considered in several works including 
\textit{Preemptive}-$TPN $  \cite{Bucci} \cite{Nigro2012} \cite{Abd2006}, \textit{Stopwatch-} $TPN$ \cite{Bert-grid}, 
\textit{Inhibitor-}$TPN$ \cite{IHTPN}, \textit{Scheduling-}$TPN$
\cite{LIme2003} and of unfolding safe parametric stopwatch TPN (PSwPNs)\cite{jard2013}. For example, in \cite{IHTPN} the authors defined the 
\textit{ITPN} (\textit{Inhibitor arc Time Petri Nets) }model, wherein the
progression and the suspension of time is driven by using standard and
inhibitor arcs.

However, whatever the model we consider, the time analysis of the system is
basically the same, as it involves the investigation of a part of or the
whole set of its reachable states that determines its state space. As the
state space is generally infinite due to dense time semantics, we need
therefore to compute finite abstractions of it, that preserve properties of
interest. In these abstractions, states are grouped together, in order to
obtain a finite number of these groups. These groups of states are, for
instance, regions and zones for timed automata, or state classes \cite%
{BerDiaz} for time Petri nets. Hence, the states pertaining to each group
can be described by a system of linear inequalities, noted $D$, whose set of
solutions determines the state space of the group. Hence, if the model does
not use any stopwatch, then $D$ is of a particular form, called \textit{DBM} 
\textit{(Difference Bound Matrix) }\cite{Dill}. However, when using
stopwatches, the system $D$ becomes more complex and does not fit anymore
into a \textit{DBM}. \ In actual fact, $D$ takes a general polyhedral form
whose \textit{canonical} form \cite{Avis} is given as a conjunction of two
subsystems $D=\overrightarrow{D}\wedge \widehat{D},$\ where $\overrightarrow{%
D}$ is a \textit{DBM}\ system\ and $\widehat{D}$ is a polyhedral system that
cannot be encoded with \textit{DBMs}.

The major shortcoming of manipulating polyhedra is the performance loss in
terms of computation speed and memory usage. Indeed, the complexity of
solving a general polyhedral system is exponential in the worst case, while
it is polynomial for a \textit{DBM} system. Furthermore, the reachability is
proved to be undecidable for both $\mathit{SWA}$ and $\mathit{LHA}$ \cite%
{Cassez} \cite{Alur} \cite{Henz}, as well as for $TPN$\ extended with
stopwatches \cite{Bert-grid} \cite{magnin2009}. As a consequence, the finiteness of the exact
state class graph construction cannot be guaranteed even when the net is
bounded.

In order to speed up the graph computation, an idea is to leave out the
subsystem $\widehat{D},$ to keep only the system $\overrightarrow{D}$ thus
overapproximating the space of $D$ to the \textit{DBM} containing it, see 
\cite{Bucci}\cite{IHTPN}\cite{abdelli} for details. The obvious consequence
of the overapproximation is that we add states in the computed group that
are not reachable indeed. Yet more, this could prevent the graph computation
to terminate, by making the number of computed markings unbounded.
Conversely, this can also make the computation of the approximated graph
terminate by cutting off the polyhedral inequalities that prevent the
convergence.

Furthermore, in order to settle a compromise between both techniques, a
hybrid approach has been proposed by \textit{Roux et al }\cite{ROUX Magnat}.
The latter puts forward a sufficient condition that determines the cases
where the subsystem $\widehat{D}$ becomes redundant in $D$. Hence, the
combination of both \textit{DBM} and polyhedral representations makes it
possible to build the exact state class graph faster and with lower expenses
in terms of memory usage comparatively to the polyhedra based approach \cite%
{LIme2003}. More recently, \textit{Berthomieu et al} have proposed an
overapproximation method based on a quantization of the polyhedral system $D$
\cite{Bert-grid}. The latter approach ends in the exact computation of the
graph in almost all cases faster than the hybrid approach \cite{ROUX Magnat}%
. Nevertheless, this technique is more costly in terms of computation time
and memory usage comparatively to the \textit{DBM} overapproximation
although it yields much precise graphs.

Different algorithms \cite{abdelli}\cite{IHTPN}\cite{Bucci} have been
defined in the literature to compute the \textit{DBM} overapproximation\ of
a class. All these approaches are assumed theoretically to compute the
tightest \textit{DBM} approximation of $D$. However, we have shown in \cite%
{abdelli} that by avoiding to compute the minimal form of the \textit{DBM}
systems, our algorithm succeeds to compute straightforwardly the reachable
systems in their normal form. We thereby shunned the computation and the
manipulation of the intermediary polyhedra. Moreover, the effort needed for
the normalization and the minimization of the resulted \textit{DBM}\ system
is removed. This has improved greatly the implementation and the computation
of the \textit{DBM} overapproximated graph.

Although the cost of computing the \textit{DBM} overapproximation is low
comparing to the exact construction, it remains that in certain cases the
approximation is too coarse to restore properties of interest and especially
quantitative properties \cite{Abdelli un}. In actual fact, more the
approximated graphs are big more the approximation looses its precision and
therefore includes false behaviors that may skew the time analysis of the
system. Many of these false behaviors are generated in the \textit{DBM}
overapproximation because the computation of a \textit{DBM} class is
performed recursively only from its direct predecessor class. We think that
some time information that stand in upper classes in the firing sequence
could be used to fix the approximation of the class to compute. In actual
fact, the \textit{DBM} overapproximations defined in \cite{abdelli}\cite%
{IHTPN}\cite{Bucci} are assumed to be\textit{\ }the tightest when referring
to the polyhhedral system $D$ computed in the context of the approximated
graph. The latter may not be equal to the polyhedral system resulted after
firing the same sequence in the exact graph. As polyhedral constraints are
removed systematically each time they appear in upper classes in the firing
sequence, the resulted $DBM$ overapproximation looses its precision.
Therefore, the \textit{DBM} overapproximation could be still more tightened
if we could restore some time information encoded by polyhedral constraints
removed in the upper classes in the firing sequence.

We explore in this paper a novel approach to compute a more precise \textit{%
DBM} overapproximation of the state space of real time preemptive systems
modeled by using the $\mathit{ITPN}$\ model. For this effect, we extend the
expression of a class to the time distance system that encodes the
quantitative properties of firing's subsequences. The time distance system has been already considered 
in the computation of the state space of many timed Petri nets extensions as \cite{Bouch} \cite{TSPN}.
This system records relevant time information that is exploited to tighten still more the 
\textit{DBM} overapproximation of a class. Although, the cost of computing
the latter is slightly higher than when using classical \textit{DBM}
overapproximation techniques \cite{abdelli}\cite{IHTPN}\cite{Bucci}, the
global effort needed to compute the final \textit{DBM} system remains
polynomial. Consequently, the resulted approximated graphs are very compact,
even equal to the exact ones while improving by far their calculation times.
Moreover, the obtained graphs are more suitable to restore quantitative
properties of the model than other constructions. To advocate the benefits
of this graph approximation, we report some experimental results comparing
our graph constructions with other fellow approaches.

The remainder of this paper is organized as follows: In section 2, we
present the syntax and the formal semantics of the $ITPN$ model. In section\
3, we lay down and discuss through an example the algorithms that build the
exact graph and the \textit{DBM} overapproximation of an $ITPN$. In section
4, we introduce formally our overapproximation and show how the approximated
graph is built. In $Section\ 5,$ we report the experimentation results of
the implementation of our algorithms and compare them with those of other
graph constructions.

\section{Time Petri Net with Inhibitor Arcs}

\textit{Time Petri nets with inhibitor arcs} ($ITPN$) \cite{IHTPN} extends
time Petri nets\cite{Merlin} to \textit{Stopwatch inhibitor arcs}. Formally,
an $ITPN$ is defined as follows:

\begin{definition}
{\small {An $\mathit{ITPN}$\textit{\ }is given by the tuple $%
(P,T,B,F,M^{0},I,IH)$ where: $P$ and $T$ are respectively two nonempty sets
of places and transitions; $B$ is the backward incidence function \footnote{%
{\small $%
%TCIMACRO{\U{2115} }%
%BeginExpansion
\mathbb{N}
%EndExpansion
$ denotes the set of positive integers. In the graphical representation, we
represent only arcs of non null valuation, and those valued 1 are implicit.}}
: $B:P\times T\longrightarrow 
%TCIMACRO{\U{2115} }%
%BeginExpansion
\mathbb{N}
%EndExpansion
=\{0,1,2,..\};$ $F$ is the forward incidence function $F:P\times
T\longrightarrow 
%TCIMACRO{\U{2115} }%
%BeginExpansion
\mathbb{N}
%EndExpansion
$ $;$ $M^{0}$ is the initial marking mapping $M^{0}:P\longrightarrow 
%TCIMACRO{\U{2115} }%
%BeginExpansion
\mathbb{N}
%EndExpansion
$ ; $I$ is the delay mapping $I:T\longrightarrow 
%TCIMACRO{\U{211a} }%
%BeginExpansion
\mathbb{Q}
%EndExpansion
^{+}\times 
%TCIMACRO{\U{211a} }%
%BeginExpansion
\mathbb{Q}
%EndExpansion
^{+}\cup \left\{ \infty \right\} ,$\ where $%
%TCIMACRO{\U{211a} }%
%BeginExpansion
\mathbb{Q}
%EndExpansion
^{+}$ is the set of non negative rational numbers. We write $%
I(t)=[tmin(t),tmax(t)]$\ such that $0\leq tmin(t)\leq tmax(t)$ $;$ $%
IH:P\times T\longrightarrow 
%TCIMACRO{\U{2115} }%
%BeginExpansion
\mathbb{N}
%EndExpansion
$ \ is the inhibitor arc function; there is an inhibitor arc connecting the
place $p$ to the transition $t,$ if $IH(p,t)\neq 0.$ }}
\end{definition}

\begin{figure}[h]
\centering\includegraphics[width=7 cm]{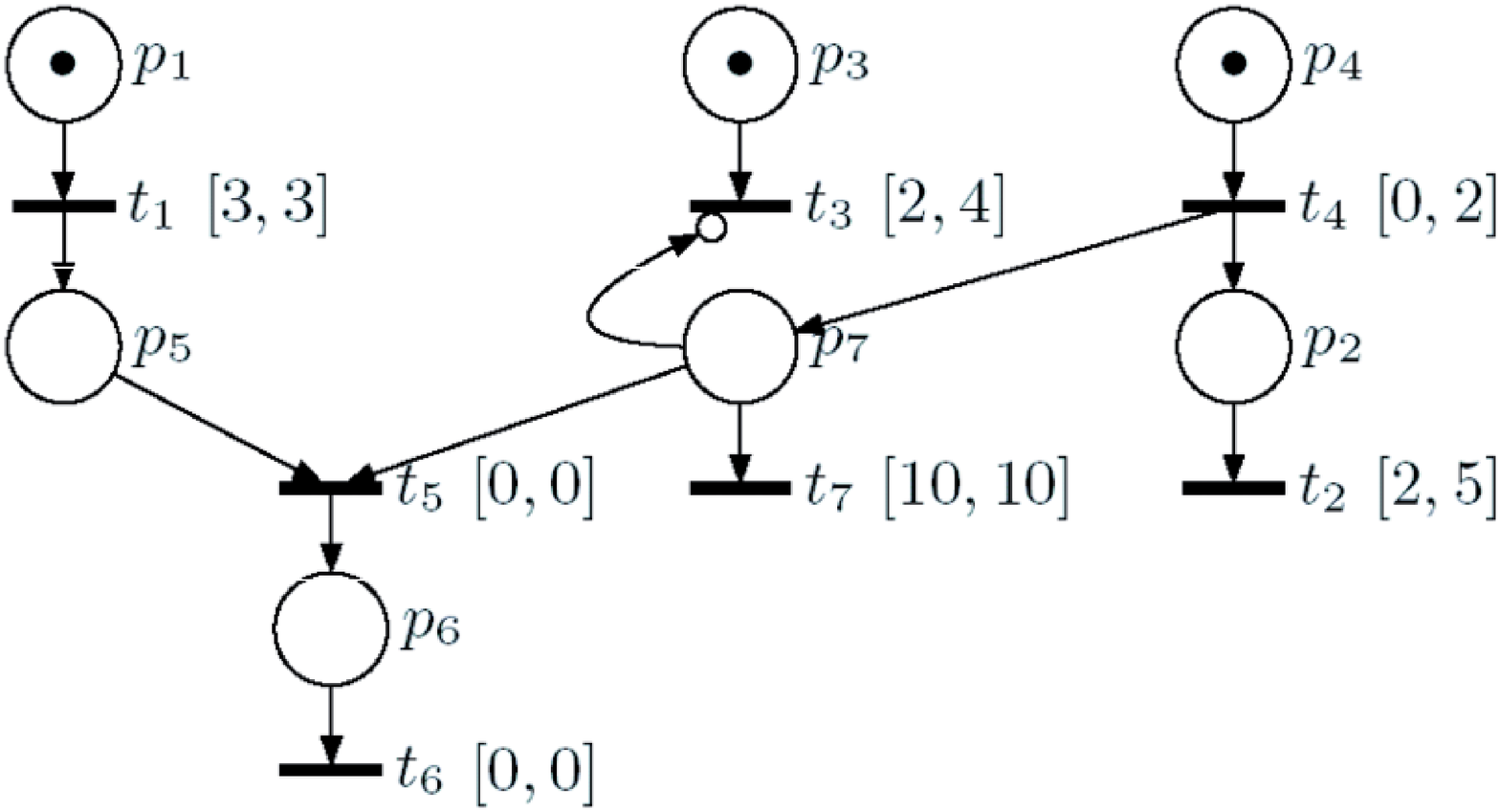}
\caption{An $ITPN$ model }
\label{Fig1}
\end{figure}

For instance, let us consider the $ITPN$ model shown in \textit{Fig 1}.
Therein, the \textit{inhibitor arc} is the arc ended by a circle that
connects the place $p_{7}$ to the transition $t_{3}$.\ Initially, the place $%
p_{3}$ is marked but the place $p_{7}$ is not; hence $t_{3}$ is enabled but
not inhibited. Therefore, $t_{3}$ is progressing as it is the case for $%
t_{4} $\ which is also enabled for the initial marking. However, the firing
of the transition $t_{4}$\ consumes the token in the place $p_{4}$ and
produces another in $p_{2}$\ and another one in $p_{7}$.\ Therefore, the
inhibitor arc becomes activated and the clock of $t_{3}$ is thus suspended ($%
t_{3}$ is inhibited). This suspension lasts as long as $p_{7}$\ remains
marked. For more details, the formal semantics of the $ITPN$ model is
introduced hereafter.

Let $RT:=(P,T,B,F,M^{0},I,IH)$ be an \textit{ITPN. }

\begin{description}
\item[-] We call a \textit{marking} the mapping, noted $M,$ which associates
with each place a number of tokens: $M:P\rightarrow 
%TCIMACRO{\U{2115} }%
%BeginExpansion
\mathbb{N}
%EndExpansion
.$

\item[-] A transition $t$ is said to be \textit{enabled} for the marking $M,$
if $\forall p\in P,B(p,t)\leq M(p)$; the number of tokens in each input
place of $t$ is greater or equal to the valuation of the arc connecting this
place to the transition $t$. Thereafter, we denote by $Te(M)$ the set of
transitions \textit{enabled} for the marking $M$.

\item[-] A transition $t$ is said to be \textit{inhibited }for a marking $M,$
if it is \textit{enabled} and if there exists an inhibitor arc connected to $%
t,$ such that the marking satisfies its valuation ($t\in Te(M))\wedge
\exists p\in P,0<IH(p,t)\leq M(p)$. We denote by $Ti(M)$ the set of
transitions that are \textit{inhibited} for the marking $M$.

\item[-] A transition $t$ is said to be \textit{activated }for a marking $M,$
if it is enabled and not inhibited, ($t\in Te(M))\wedge $ $\ (t\notin Ti(M))$%
; we denote by $Ta(M)$ the set of transitions that are \textit{activated}
for the marking $M$.

\item[-] Let $M$ be a marking ; two transitions $t_{i}$ and $t_{j}$ enabled
for $M$ are said to be \textit{conflicting} for $M$, if $\exists p\in
P,\quad B(p,t_{i})+B(p,t_{j})>M(p).$

\item[-] We note hereafter by $Conf(M)$ the relation built on $Te(M)^{2}$
such that $(t_{1},t_{2})\in Conf(M),$ iff $t_{1}$ and $t_{2}$ are in
conflict for the marking $M$.
\end{description}

For instance, let us consider again the \textit{$ITPN$} of \textit{Fig 1}.
Its initial marking is equal to $M^{0}:\left\{ p_{1},p_{3},p_{4}\right\}
\rightarrow 1;\left\{ p_{2},p_{5},p_{6},p_{7}\right\} \rightarrow 0.$ the
sets of enabled, inhibited, and activated transitions for $M^{0}$\ are
respectively $Te(M^{0})=\left\{ t_{1},t_{3,}t_{4}\right\} ,$\ $%
Ti(M^{0})=\varnothing ,$\ and $Ta(M^{0})=Te(M^{0}).\medskip $

\begin{remark}
We assume in the sequel a monoserver semantics, which means that no
transition can be enabled more than once for any marking.
\end{remark}

We define the semantics of an $ITPN$ as follows:

\begin{definition}
{\small {\ The semantics of an $ITPN$ is defined as a LTS (labeled
transition system), $ST=(\Gamma ,e^{0},\rightarrow ),$ such that: }}

\begin{itemize}
\item {\small $\Gamma $ is the set of reachable states: Each state, noted $%
e, $ pertaining to $\Gamma $ is a pair $(M,V)$ where $M$\ is a marking and $%
V $\ is a valuation function that associates with each enabled transition $t$
of $Te(M)$ a time interval that gives the range of relative times within
which $t $ can be fired. Formally we have : $\forall t$\ $\in Te(M),\quad
V(t):=[x(t),y(t)]$ }

\item {\small $e^{0}=(M^{0},V^{0})$ is the initial state, such that: $%
\forall t\in Te(M^{0}),\quad V^{0}(t):=I(t):=[tmin(t),tmax(t)].$ }

\item {\small $\rightarrow \in \Gamma \times (T\times 
%TCIMACRO{\U{211a} }%
%BeginExpansion
\mathbb{Q}
%EndExpansion
^{+})\times \Gamma $ \ is a relation, such that $((M,V),(t_{f},\underline{%
t_{f}}),(M^{\uparrow },V^{\uparrow }))\in \rightarrow ,$ iff: }

\begin{description}
\item[(i)] {\small $t_{f}\in Ta(M).$ }

\item[(ii)] {\small $x(t_{f})\leq \underline{t_{f}}\leq \underset{\forall
t\in Ta(M)}{MIN}\left\{ y(t)\right\} .$ }
\end{description}

{\small and we have: }

{\small $\forall p\in P,$ $M^{\uparrow
}(p):=M(p)-B(p,t_{f})+F(p,t_{f})\smallskip .$ }

{\small $\forall t\in Te(M^{\uparrow })$ }

{\small if $t$ $\notin New(M^{\uparrow })$: }

{\small $\quad 
\begin{tabular}{ll}
$\lbrack x^{\uparrow }(t),\ y^{\uparrow }(t)]:=[MAX(0,\ x(t)-\underline{t_{f}%
}),$ $\ y(t)-\underline{t_{f}}]$ & $t\in Ta(M)$ \\ 
$\lbrack x^{\uparrow }(t),\ y^{\uparrow }(t)]:=[x(t),$ $\ y(t)]$ & $t\in
Ti(M)$%
\end{tabular}%
\medskip $ }

{\small if $t$ $\in New(M^{\uparrow })$\quad }

{\small $\quad \lbrack x^{\uparrow }(t),\ y^{\uparrow }(t)]:=I(t)=[tmin(t),\
tmax(t)]$ \smallskip }

\begin{itemize}
\item {\small where $New(M^{\uparrow })$ denotes the set of transitions 
\textit{newly enabled} for the marking $M^{\uparrow }.$ These transitions
are those enabled for $M^{\uparrow }$\ and not for $M$, or those enabled for 
$M^{\uparrow }$\ and $M$ but are conflicting with $t_{f}$ \ for the marking $%
M$. \ Otherwise, an enabled transition which does not belong to $%
New(M^{\uparrow })$ is said to be \textit{persistent}. }
\end{itemize}
\end{itemize}
\end{definition}

If $t$ is a transition enabled for the state $e$, we note \underline{$t$}
the clock associated with $t$ that takes its values in $%
%TCIMACRO{\U{211a} }%
%BeginExpansion
\mathbb{Q}
%EndExpansion
^{+}.$ \underline{$t$} measures the residual time of the transition $t$
relatively to the instant where the state $e$ is reached. The time
progresses only for activated transitions, whereas it is suspended for
inhibited transitions. Therefore, a transition $t_{f}$ can be fired at
relative time $\underline{t_{f}}$ from a reachable state $e,$ if $(i)$ $%
t_{f} $ is \textit{activated} for the marking $M$, and if $(ii)$ the time
can progress within the firing interval of $t_{f}$ without overtaking those
of other activated transitions. After firing $t_{f}$ the reachable state,
noted $e^{\uparrow },$ is obtained:

\begin{itemize}
\item by consuming a number of tokens in each input place $p$ of $t_{f}$
(given by the value $B(p,t_{f})$), and by producing a number of tokens in
each output place $p$ of $t_{f}$ (given by the value $F(p,t_{f})$);

\item by shifting the interval of a persistent activated transition with the
value of the firing time of $t_{f}$. However, the intervals of persistent
inhibited transitions remain unchanged. Finally, a newly enabled transition
is assigned its static firing interval.
\end{itemize}

Similarly as for a $TPN,$ the behavior of an $ITPN$\ can be defined as a
sequence of pairs $(t_{f}^{i},\underline{t_{f}^{i}})$, where $t_{f}^{i}$ is
a transition of the net and $\underline{t_{f}^{i}}$ $\in 
%TCIMACRO{\U{211a} }%
%BeginExpansion
\mathbb{Q}
%EndExpansion
^{+}$. Therefore, the sequence $\mathit{S}^{\mathit{\ast }}=((t_{f}^{1},%
\underline{t_{f}^{1}}),(t_{f}^{2},\underline{t_{f}^{2}}),..,(t_{f}^{n},%
\underline{t_{f}^{n}}))$\ denotes that $t_{f}^{1}$\ is firable after $%
\underline{t_{f}^{1}}$\ time units, then $t_{f}^{2}$\ is fired after $%
\underline{t_{f}^{2}}$\ time units and so on, such that $t_{f}^{n}$ is fired
after the absolute time $\sum_{i=1}^{n}\underline{t_{f}^{i}}.$ Moreover, we
often express the behavior of the net as an \textit{untimed sequence, }%
denoted by\textit{\ }$\mathit{S}$\textit{,} obtained from a timed sequence $%
\mathit{S}^{\mathit{\ast }}$ by removing the firing times: If $\mathit{S}^{%
\mathit{\ast }}=((t_{f}^{1},\underline{t_{f}^{1}}),(t_{f}^{2},\underline{%
t_{f}^{2}}),..,(t_{f}^{n},\underline{t_{f}^{n}})),$ then $\mathit{S}%
=(t_{f}^{1},t_{f}^{2},..,t_{f}^{n}).$ As the set of time values is assumed
to be dense, the model $ST$ is infinite. In order to analyze this model, we
need to compute an abstraction of it that saves the most properties of
interest. The construction of a \textit{symbolic graph} preserves the
untimed sequences of $ST,$ and makes it possible to compute a finite graph
in almost all cases. We show hereafter how to compute the state class graph
of the $ITPN$ that preserves chiefly the linear properties of the model.

\section{$ITPN$ state space construction}

As for a $TPN$ model \cite{Merlin}, the state graph $ST$ of an $ITPN$ can be
contracted by gathering in a same class all the states reachable after
firing the same untimed sequence. This approach (known as the state class
graph method \cite{BerDiaz}), expresses each class as a pair $(M,D)$ where $%
M $ is is the common marking and $D$ is a system of inequalities that
encodes the state space of the class. Each variable of such a system is
associated with an enabled transition and measures its residual time$.$ When
dealing with an $ITPN$, the inequalities of the system $D$ may take a
polyhedral form \cite{LIme2003}. More formally, a class of states of an $%
ITPN $\ is defined as follows:

\begin{definition}
{\small {\ Let $ST=(\Gamma ,e^{0},\rightarrow )$ be the LTS associated with
an $ITPN$. A class of states of an $ITPN$, denoted by $E,$\ is the set of
all the states pertaining to $\Gamma $ that are reachable after firing the
same untimed sequence $S=$ $(t_{f}^{1},..,t_{f}^{n})$ from the initial state 
$e^{0}$. A class $E$\ is defined by $(M,D),$ where $M$ is the marking
reachable after firing $S$, and $D$ is the firing space encoded as a set of
inequalities. \newline
For $Te(M)=\{t_{1},..,t_{s}\},$ we have : \medskip $\ D=\widehat{D}\wedge $}}%
$\overrightarrow{{\small {D}}}${\small {\ }}

$\overrightarrow{{\small D}}${\small $:=\left\{ 
\begin{tabular}{l}
{\Huge $\wedge $}$_{i\neq j}$ $\quad (\underline{t_{j}}-\underline{t_{i}}%
~\leq ~d_{ij})$ \\ 
{\Huge $\wedge $}$_{i\leq s}\quad (d_{i\bullet }\leq ~\underline{t_{i}}~\leq
d_{\bullet i})$%
\end{tabular}%
\right. ~~\smallskip $ }

{\small \qquad with ($t_{j},t_{i})\in Te(M)^{2}~$ $d_{ij}\in 
%TCIMACRO{\U{211a} }%
%BeginExpansion
\mathbb{Q}
%EndExpansion
\cup \left\{ \infty \right\} ,~d_{\bullet i}\in 
%TCIMACRO{\U{211a} }%
%BeginExpansion
\mathbb{Q}
%EndExpansion
^{+}\medskip \cup \left\{ \infty \right\} ,d_{i\bullet }\in 
%TCIMACRO{\U{211a} }%
%BeginExpansion
\mathbb{Q}
%EndExpansion
^{+}$ }

{\small $\widehat{D}:=${\Huge $\wedge $ }$_{k=1..p}\quad (\alpha _{1k}%
\underline{t_{1}}+..+\alpha _{sk}\underline{t_{s}}~\leq ~d_{k})~\ $ }

{\small \qquad with$\ d_{k}\in 
%TCIMACRO{\U{211a} }%
%BeginExpansion
\mathbb{Q}
%EndExpansion
\cup \left\{ \infty \right\} ,$ $(\alpha _{1k},..,\alpha _{sk})\in 
%TCIMACRO{\U{2124} }%
%BeginExpansion
\mathbb{Z}
%EndExpansion
^{s}$ \ and\footnote{{\small $Z$ denotes the set of relative integers.\ }} }

{\small \qquad \qquad \qquad $\forall k,\exists (i,j),(\alpha _{ik},\alpha
_{jk})\notin \left\{ (0,0),(1,-1),(1,0),(-1,0)\right\} $ }
\end{definition}

We denote by the element $\left\{ \bullet \right\} $\ the instant at which
the class $E$ is reached. Therefore, the value of the clock \underline{$%
t_{i} $} expresses the time relative to the instant $\bullet ,$ at which the
transition $t_{i}$ can be fired.\ Thus, for each valuation $\psi $
satisfying the system $D,$ it corresponds a unique state $e=(M,V)$ reachable
in $ST$ after firing the sequence $S$.

In case of a $\mathit{TPN}$, the system $D$ is reduced to the subsystem $%
\overrightarrow{D}.$ The inequalities of the latter have a particular form,
called $DBM$ (\textit{Difference Bound Matrix})\cite{Dill}. The
coefficients, $d_{\bullet i},d_{i\bullet }$ and $d_{ij}$ are respectively,
the minimum residual time to fire the transition $t_{i},$ the maximum
residual time to fire the transition $t_{i},$ and the maximal firing
distance of the transition $t_{j}$ relatively to $t_{i}.$ The $DBM$ form
makes it possible to apply an efficient algorithm to compute a class, whose
overall complexity is $O(m^{3})$, where $m$\ is the number of enabled
transitions. However, for $TPN$ augmented with stopwatches, the state space
of a class cannot be encoded only with $DBMs$. Actually, inequalities of
general form (called also \textit{polyhedra)}$,$ are needed to encode this
space. The manipulation of these constraints, given by the subsystem $%
\widehat{D},$ induces a higher complexity that can be exponential in the
worst case.

The exact state class graph, noted $GR,$ of an $ITPN$\ is computed by
enumerating all the classes reachable from the initial class $E^{0}$ until
it remains no more class to explore. Formally, the exact state class graph
of an $ITPN$ can be defined as follows \cite{Bert-grid}:

\begin{definition}
{\small {The exact state class graph of an $ITPN$, denoted by $GR$, is the
tuple $(CE,E^{0},\longmapsto )$\ where: \newline
- $CE$\ is the set of classes reachable in $GR;$\ \newline
- $E^{0}=(M^{0},D^{0})$\ is the initial class such that: $D^{0}=\left\{ 
\begin{tabular}{ll}
$\forall t_{i}\in Te(M^{0}),$ & $tmin(t_{i})\leq \underline{t_{i}}\leq
tmax(t_{i})$%
\end{tabular}%
\right. $; \newline
- $\longmapsto $\ is the transition relation between classes defined on $%
CE\times T\times CE,$\ such that \newline
$((M,D),t_{f},(M^{\uparrow },D^{\uparrow }))\in \longmapsto ,$ iff: }}

\begin{description}
\item[a)] {\small $t_{f}$ is activated and the system $D$ augmented with the
firing constraints of $t_{f}$\ that we write $D_{a}=$ $D\wedge (\forall t\in
Ta(M),\quad \underline{t_{^{f}}}\leq \underline{t})$ holds. }

\item[b)] {\small $\forall p\in P,$ $M^{\uparrow
}(p):=M(p)-B(p,t_{f})+F(p,t_{f})\smallskip .$ }

\item[c)] {\small The system $D^{\uparrow }$ is computed from $D,$ as
follows: }

\begin{enumerate}
\item {\small In the system $D_{a}$, replace each variable \underline{$t$}
related to a persistent transition activated for $M$ by: $\underline{t}:=%
\underline{t_{f}}+\underline{t^{\prime }},$ thus denoting the time
progression. On the other hand, replace each variable \underline{$t$}
related to a persistent transition inhibited for $M$ by: $\underline{t}:=%
\underline{t^{\prime }},$ thus denoting the time inhibition. }

\item {\small Eliminate then by substitution the variable \underline{$t_{f}$}
as well as all the variables relative to transitions disabled by the firing
of $t_{f};$ }

\item {\small Add to the system thus computed, the time constraints relative
to each newly enabled transition for $M^{\uparrow }$: 
\begin{tabular}{ll}
$\forall t_{i}\in New(M^{\uparrow }),$ & $tmin(t_{i})\leq \underline{t_{i}}$$%
\leq tmax(t_{i})$%
\end{tabular}
}
\end{enumerate}
\end{description}
\end{definition}

The last definition shows how the exact state class graph of an $ITPN$ is
built. Being given a class $E=(M,D)$ and a transition $t_{f}$ activated for $%
M$, the computation of a class $E^{\uparrow }=(M^{\uparrow },D^{\uparrow })$%
\ \ reachable from $E$ by firing $t_{f}$ consists in computing the reachable
marking $M^{\uparrow }$ and the system $D^{\uparrow }$\ that encodes the
firing space of $E^{\uparrow }.$ The class $E$ can fire the activated
transition $t_{f},$ if there exists a valuation that satisfies $D$ (a state
of $E$), such that $t_{f}$\ can be fired before all the other activated
transitions. The firing of $t_{f}$ produces a new class $E^{\uparrow
}=(M^{\uparrow },D^{\uparrow });$ the latter gathers all the states
reachable from those of $E$ that satisfy the firing condition of $Definition$
$2.$ The system $D^{\uparrow }$ that encodes the space of $E^{\uparrow }$\
is computed from the system $D$\ augmented with the firing constraints of $%
t_{f}$. The substitution of variables relative to activated transitions
allows to shift the time origin towards the instant at which the new class $%
E^{\uparrow }$ is reached. Then, an equivalent system is computed wherein
the variables relative to transitions that have been disabled following the
firing of $t_{f}$ are removed. Finally, the constraints of transitions newly
enabled are added.

The complexity of the firing test and the step 2 of the previous algorithm
depends on the form of the system $D$. If $D$\ includes polyhedral
constraints, then the complexity of the algorithm is exponential, whereas it
is polynomial otherwise.\ It should be noticed that the system $D^{0}$
related to the initial class is always in $DBM$ form, and that polyhedral
constraints are generated in the systems of reachable classes only when both
inhibited and activated transitions stand persistently enabled in a firing
sequence \cite{Bucci} \cite{ROUX Magnat}.

Knowing how to compute the successors of a class, the state class graph
computation is basically a depth-first or breadth-first graph generation.
Then the state class graph is given as the quotient of \ $GR$ by a suitable\
equivalence relation. This equivalence relation may be equality : two
classes $(M,D)$ and $(M,D^{\prime }),$ given in their minimal form are equal
if \ $D=D^{\prime }$, or inclusion; in other terms, if $\left\rceil
D\right\lceil $\ denotes the set of solutions for the system $D$, then we
have : $\left\rceil D\right\lceil \subseteq \left\rceil D^{\prime
}\right\lceil .$ It should be noticed that the equality preserves mainly the
untimed language of the model, whereas the inclusion preserves the set of
reachable markings.

The algorithm given in \textit{Definition 4} can be applied to a $TPN$ with
the specificity that the system $D$ is always encoded in $DBM$. Moreover, it
is proved that the number of equivalent $DBM$ systems computed in the graph
is always finite \cite{BerDiaz}. This property is important since it implies
that the graph is necessarily finite, if the number of reachable markings is
bounded. Unfortunately, this last property is no more guaranteed in presence
of stopwatches. In actual fact, the number of reachable polyhedral systems
may be infinite too, thus preventing the termination of the graph
construction even when the net is bounded. To tackle these issues, the $%
\mathit{DBM}$ overapproximation technique has been proposed as an
alternative solution to analyze preemptive real time systems \cite{abdelli}%
\cite{IHTPN}\cite{Bucci}. This approach consists in cutting off the
inequalities of the subsystem $\widehat{D}$\ when the latter appears in $D$.
It thereby keeps only those of the subsystem $\overrightarrow{D}$ to
represent an overapproximation of the space of $D$. This solution makes it
possible to build a less richer graph than the exact one, but nevertheless
with lesser expenses in terms of computation time and memory usage.
Moreover, this overapproximation ensures that the number of $DBM$ systems to
be considered in the computation is always finite, whereas that of polyhedra
systems may be infinite. This may thus make the overapproximated
construction terminate, while the exact one does not.\ To better understand
how works this approach, we apply the state class graph method to the $ITPN$
example of \textit{Fig 1}. In the sequel, we denote by\ $\widetilde{%
{\scriptsize {\normalsize D}}}$ the $DBM$ system obtained by $DBM$
overapproximation which may be different from $\overrightarrow{D}$ as we can
see thereafter. Therefore, the system $\overrightarrow{D}$ denotes the
tightest $DBM$ system that one can obtain by $DBM$ overapproximation.

Let $E=(M,D)$ be the class reachable in the exact graph after firing the
sequence $(t_{4},t_{1})$ from the initial class $E^{0}=(M^{0},D^{0}).$

{\scriptsize {\textit{E}$^{0}$\textit{=}$\left( 
\begin{tabular}{l}
${\normalsize M}^{0}{\normalsize :p}_{1}{\normalsize ,p}_{3}{\normalsize ,p}%
_{4}{\normalsize \rightarrow 1}$ \\ 
${\normalsize D}^{0}{\normalsize :}\left\{ 
\begin{tabular}{l}
3 $\leq $ $\underline{t_{1}}\leq $ 3 \\ 
2 $\leq $ $\underline{t_{3}}\leq $ 4 \\ 
0 $\leq \underline{t_{4}}\leq $ 2%
\end{tabular}%
\right. $%
\end{tabular}%
\right. $ \ \textit{E=}$\left( 
\begin{tabular}{l}
${\normalsize M:p}_{2}{\normalsize ,p}_{3}{\normalsize ,p}_{5}{\normalsize ,p%
}_{7}{\normalsize \rightarrow 1}$ \\ 
${\normalsize D:}\left\{ 
\begin{tabular}{l}
$\underline{t_{5}}=0$ \\ 
$-8\leq \underline{t_{2}}-\underline{t_{7}}\leq -5$ \\ 
$7\leq \underline{t_{7}}\leq 9$ \\ 
$0\leq \underline{t_{2}}$ \\ 
$9\leq \underline{t_{7}}+\underline{t_{3}}\leq 11$%
\end{tabular}%
\right. $%
\end{tabular}%
\right. $ $\ \widetilde{{\normalsize D}}{\normalsize :}\left\{ 
\begin{tabular}{l}
$\underline{t_{5}}=0$ \\ 
$-8\leq \underline{t_{2}}-\underline{t_{7}}\leq -5$ \\ 
$7\leq \underline{t_{7}}\leq 9$ \\ 
$0\leq \underline{t_{2}}$ \\ 
$0\leq \underline{t_{3}}\leq 4$%
\end{tabular}%
\right. $ } }

At this stage, polyhedral constraints given by $9\leq \underline{t_{7}}+%
\underline{t_{3}}\leq 11$ appear for the first time in the firing sequence.
This happens because the inhibited transition $t_{3}$ and the activated
transitions $t_{7}$ and $t_{2}$ are persistently enabled in this sequence.
The $\mathit{DBM}$ overapproximation consists in cutting off the polyhedral
constraints $9\leq \underline{t_{7}}+\underline{t_{3}}\leq 11$ after
normalizing all the $DBM$ constraints. We thereby obtain the system $%
\widetilde{{\scriptsize {\normalsize D}}}$ that replaces the system $%
{\normalsize D}$ in the $DBM$ approximated class $\widetilde{{\normalsize E}}
$. However, at this stage, the removed polyhedral constraints are redundant
relatively to $\widetilde{{\scriptsize {\normalsize D}}}$ and therefore have
no impact on the firing of activated transitions $t_{2}$, $t_{5}$ and $t_{7}$%
. Let us consider now the firing of the transition $t_{2}$ from both classes 
${\normalsize E}$ and $\widetilde{{\normalsize E}}$ to reach respectively
the classes ${\normalsize E}^{\prime }$ and $\widetilde{{\normalsize E}%
^{\prime }}.$

{\scriptsize {\ \textit{E}$^{\prime }$\textit{=}$\left( 
\begin{tabular}{l}
${\normalsize M}^{\prime }{\normalsize :p}_{3}{\normalsize ,p}_{5}%
{\normalsize ,p}_{7}{\normalsize \rightarrow 1}$ \\ 
${\normalsize D}^{\prime }{\normalsize :}\left\{ 
\begin{tabular}{l}
$\underline{t_{5}}=0$ \\ 
$7\leq \underline{t_{7}}\leq 8$ \\ 
$9\leq \underline{t_{7}}+\underline{t_{3}}\leq 11$%
\end{tabular}%
\right. $%
\end{tabular}%
\right. $ $\ \widetilde{\mathit{E}^{\prime }}$\textit{=}$\left( 
\begin{tabular}{l}
${\normalsize M}^{\prime }{\normalsize :p}_{3}{\normalsize ,p}_{5}%
{\normalsize ,p}_{7}{\normalsize \rightarrow 1}$ \\ 
$\widetilde{{\normalsize D}^{\prime }}{\normalsize :}\left\{ 
\begin{tabular}{l}
$\underline{t_{5}}=0$ \\ 
$-8\leq \underline{t_{5}}-\underline{t_{7}}\leq -7$ \\ 
$7\leq \underline{t_{7}}\leq 8$ \\ 
$0\leq \underline{t_{3}}\leq 4$%
\end{tabular}%
\right. $%
\end{tabular}%
\right. $ } }

As we notice, the polyhedral constraints are still present in $E^{\prime }$
since the transitions $t_{3}$ and $t_{7}$ remain persistently enabled.
However, these constraints are no more redundant relatively to the system $%
\widetilde{{\normalsize D}^{\prime }}$ as we obtain the DBM constraints $%
1\leq \underline{t_{3}}\leq 4$ in $\overrightarrow{{\normalsize D}^{\prime }}
$ after normalisation$.$ Therefore, this loss in the precision in the 
\textit{DBM} overapproximation may have an impact on the firing process
ahead in the sequence. To highlight this fact, let us consider the firing of
the transition $t_{5}$ from both classes ${\normalsize E}^{\prime }$ and $%
\widetilde{{\normalsize E}^{\prime }}$ to reach respectively the classes $%
{\normalsize E}"$ and $\widetilde{{\normalsize E}"}.$

{\scriptsize {\textit{E}$"$\textit{=}$\left( 
\begin{tabular}{l}
${\normalsize M}"{\normalsize :p}_{3}{\normalsize ,p}_{6}{\normalsize %
\rightarrow 1}$ \\ 
${\normalsize D}"{\normalsize :}\left\{ 
\begin{tabular}{l}
$\underline{t_{6}}=0$ \\ 
$1\leq \underline{t_{3}}\leq 4$%
\end{tabular}%
\right. $%
\end{tabular}%
\right. $ \ $\ \widetilde{\mathit{E}"}$\textit{=}$\left( 
\begin{tabular}{l}
${\normalsize M}"{\normalsize :p}_{3}{\normalsize ,p}_{6}{\normalsize %
\rightarrow 1}$ \\ 
$\widetilde{{\normalsize D}"}{\normalsize :}\left\{ 
\begin{tabular}{l}
$\underline{t_{6}}=0$ \\ 
$0\leq \underline{t_{3}}\leq 4$%
\end{tabular}%
\right. $%
\end{tabular}%
\right. $ } }

At this stage, we notice that both systems $D"=\overrightarrow{D"}$ and $%
\widetilde{D"}$ are both in \textit{DBM}, but the exact system $D"$is more
precise that the one obtained by overapproximation. As a result, only the
transition $t_{6}$ is firable from ${\normalsize E}"$ but not $t_{3}$ since $%
{\normalsize D}"\wedge (t_{3}\leq t_{6})$ is not consistent. However, due to
constraints relaxation both transitions are firable from $\widetilde{\mathit{%
E}"}.$ Hence we have an additional sequence in the \textit{DBM}
overapproximated graph that is not reachable in the exact graph $GR$.

In actual fact, all is about the minimal residual time of $t_{3}$\ which has
increased during its inhibition time from 0 to 1. Let us clarify this point,
initially $t_{3}$ is activated and we have $2\leq \underline{t_{3}},$ and
the model fires the transition $t_{4}$ between $[0,2]$. After this firing,
the transition $t_{2}$\ is enabled for the first time, the place $p_{7}$
becomes marked, and $t_{3}$ is inhibited for the first time; we have $0\leq $%
\underline{$t_{3}$} and $2\leq \underline{t_{2}}\leq 5.$\ The transition $%
t_{1}$ is fired afterwards to enable the transition $t_{5}$ and we have 
\underline{$t_{5}$}=0. So to be able to fire the persistent transition $%
t_{2} $, we must have (\underline{$t_{2}$}$=0)$ too. This compels the
relative time to progress at least with $tmin(t_{2})=2$ when firing $t_{1}$,
while the elapsed absolute time must not surpass $tmax(t_{1})=3$. This last
constraint restricts the state space of the class reachable after firing $%
t_{2}$\ only to states\footnote{%
For these states, the transition $t_{3}$ is not yet inhibited.} that have
fired initially $t_{4}$ during $[0,1]$. As a result, the minimal residual
time of the inhibited transition $t_{3}$ increases to 1 after the firing of $%
t_{2}$\textit{. }

The loss of precision in $\widetilde{D"}$ comparatively to $\overrightarrow{%
D"}$ is due to some polyhedral constraints involved in the normalization of $%
\overrightarrow{D"}$ that are removed in the predecessor classes of $%
\widetilde{\mathit{E}"}$. Therefore, we think that some time information
that stand in the upper classes in the firing sequence could be used to fix
the problem and to tighten still more the approximation. This will be the
subject of our proposal which is addressed in the next section. But before
we need to introduce formally the construction of the \textit{DBM}
overapproximation graph.

The computation of the $\mathit{DBM}$ overapproximation of a class $E$\ can
be obtained by using different algorithms \cite{abdelli}\cite{IHTPN}\cite%
{Bucci}. However, we have shown in a previous work \cite{abdelli} that by
avoiding to compute the $DBM$ systems systematically in their minimal form,
we succeed to define an algorithm that computes straightforwardly the
reachable systems in their normal form. We thereby shunned the computation
and the manipulation of the intermediary polyhedra. Moreover, the effort
needed for the normalization and the minimization of the resulted \textit{DBM%
}\ system is removed; this improves greatly the implementation and the
computation of the \textit{DBM} overapproximated graph. This algorithm
encodes the full $DBM$ system $\widetilde{D}$ as a square matrix where each
line and corresponding column, are indexed by an element of $Te(M)\cup
\left\{ \bullet \right\} .$ In concrete terms, we have: $\forall
(t_{i},t_{j})\in Te(M)^{2}\wedge (t_{i}\neq t_{j}),\quad \widetilde{D}%
[\bullet ,t_{i}]:=$ $d_{\bullet i};$\ $\quad \widetilde{D}[t_{i},\bullet
]:=-d_{i\bullet }$ $;$\newline
$\quad \quad \widetilde{D}[t_{i},t_{j}]:=d_{ij};\quad \widetilde{D}%
[t_{i},t_{i}]:=0$ $;\quad \quad \widetilde{D}[\bullet ,\bullet ]:=0.$

\begin{table}[h]
\caption{The matrix representation of the system $\widetilde{D^{0}}$\textit{.%
}}\centering\centering 
\begin{tabular}{|l||l|l|l|l|}
\hline
$\widetilde{D^{0}}$ & $\bullet $ & $t_{1}$ & $t_{3}$ & $t_{4}$ \\ 
\hline\hline
$\bullet $ & {\scriptsize 0} & {\scriptsize 3} & {\scriptsize 4} & 
{\scriptsize 2} \\ \hline
$t_{1}$ & {\scriptsize -3} & {\scriptsize 0} & {\scriptsize 1} & 
{\scriptsize -1} \\ \hline
$t_{3}$ & {\scriptsize -2} & {\scriptsize 1} & {\scriptsize 0} & 
{\scriptsize 0} \\ \hline
$t_{4}$ & {\scriptsize 0} & {\scriptsize 3} & {\scriptsize 4} & {\scriptsize %
0} \\ \hline
\end{tabular}%
\end{table}

These matrix notations are used to represent the coefficients of the system $%
\widetilde{D}$. For example, the matrix shown in \textit{Tab.1} encodes the
system $\widetilde{D^{0}}=D^{0}$ associated with the initial class of the $%
\mathit{ITPN}$\ of \textit{Fig 1. }

The construction of the \textit{DBM} overapproximation graph, noted $%
\widetilde{GR}$, can be computed as follows \cite{abdelli}:

\begin{definition}
{\small {\ The DBM overapproximated graph of an $ITPN$, noted $\widetilde{GR}
$, is the tuple $(\widetilde{CE},\widetilde{E^{0}},\rightsquigarrow ),$\
such that : }}

\begin{itemize}
\item {\small $\widetilde{CE}$\ is the set of DBM\ overapproximated classes
reachable in $\widetilde{GR\text{ }};$\ }

\item {\small $\widetilde{E^{0}}=(M^{0},\widetilde{D^{0}})\in \widetilde{CE}$
is the initial class, such that: }

{\small $\widetilde{D^{0}}:=\left\{ 
\begin{tabular}{ll}
$\forall t_{i}\in Te(M^{0}),$ & $tmin(t_{i})\leq \underline{t_{i}}\leq
tmax(t_{i})$ \\ 
$\forall t_{i}\neq t_{j}\in Te(M^{0}),$ & $\underline{t_{j}}-\underline{t_{i}%
}\leq tmax(t_{j})-tmin(t_{i})$%
\end{tabular}%
\right. $ }

\item {\small \ $\rightsquigarrow $ \ is a transition relation between DBM
overapproximated classes defined on $\widetilde{CE}\times T\times \widetilde{%
CE},$\ such that $((M,\widetilde{D}),t_{f},(M^{\uparrow },\widetilde{%
D^{\uparrow }}))\in \rightsquigarrow ,$ iff : }

\begin{itemize}
\item {\small $\left( t_{f}\in Ta(M)\right) $ $\wedge $ $(\widetilde{\beta }%
[t_{f}]\ \geq 0)$ such that:\ $\forall x\in Te(M)\cup \left\{ \bullet
\right\} ,\quad \widetilde{\beta }[x]=\underset{\forall t\in Ta(M)}{MIN}%
\left\{ \widetilde{D}[x,t]\right\} $. }

\item {\small $\forall p\in P,$ $M^{\uparrow
}(p):=M(p)-B(p,t_{f})+F(p,t_{f})\smallskip .$ }

\item {\small The coefficients of the $DBM$\ \ inequalities of the system $%
\widetilde{D^{\uparrow }}$\ are computed from those of $\widetilde{D}$\ by
applying the following algorithm: \bigskip }

{\small $\forall t\in Te(M^{\uparrow })$ }

{\small $\quad \quad \widetilde{D^{^{\uparrow }}}[t,t]:=0;\quad \quad \quad 
\widetilde{D^{^{\uparrow }}}[\bullet ,\bullet ]:=0;$ }

{\small \quad If $t$ is persistent }

{\small \quad \quad If $t\in Ti(M)$ $\ (t$ is inhibited for $M$) }

{\small $\quad \quad \quad \widetilde{D^{^{\uparrow }}}[t,\bullet
]:=MIN\left( 
\begin{tabular}{l}
$\widetilde{D}[t,\bullet ]$ \\ 
$\widetilde{D}[t_{f},\bullet ]+\widetilde{\beta }[t]$%
\end{tabular}%
\right. \quad \widetilde{D^{\uparrow }}[\bullet ,t]:=MIN\left( 
\begin{tabular}{l}
$\widetilde{D}[\bullet ,t]\medskip $ \\ 
$\widetilde{D}[t_{f},t]+\widetilde{\beta }[\bullet ]$%
\end{tabular}%
\right. $ \medskip }

{\small \quad \quad If $t\notin Ti(M)$ $\ (t$ is not inhibited for $M$) }

{\small $\quad \quad \quad \widetilde{D^{\uparrow }}[\bullet ,t]:=\widetilde{%
D}[t_{f},t]$ $;\quad \quad \quad \quad \widetilde{D^{^{\uparrow }}}%
[t,\bullet ]:=\medskip \widetilde{\beta }[t].$ }

{\small \quad If $t$ is newly enabled. }

{\small $\quad \quad \widetilde{D^{\uparrow }}[\bullet ,t]:=tmax(t)$ $;\quad
\quad \ \ \widetilde{D^{^{\uparrow }}}[t,\bullet ]:=-tmin(t).\medskip
\bigskip $ }

{\small $\forall (t_{1},t_{2})\in (Te(M^{\uparrow }))^{2}\wedge (t_{1}\neq
t_{2})$ }

{\small \quad If $t_{1}$ or $t_{2}$ are newly enabled. }

{\small $\ \ \quad \widetilde{D^{\uparrow }}[t_{1},t_{2}]:=\widetilde{%
D^{\uparrow }}[\bullet ,t_{2}]+\widetilde{D^{^{\uparrow }}}[t_{1},\bullet ]$%
.\medskip \medskip }

{\small \quad If $t$$_{1}$\ and $t$$_{2}$ are persistent. }

{\small $\quad $\quad If ($t_{1},t_{2})\notin (Ti(M))^{2}$\ \ ($t$$_{1}$\
and $t$$_{2}$\ are not inhibited for $M$) }

{\small $\quad \quad \quad \widetilde{D^{\uparrow }}[t_{1},t_{2}]:=MIN(%
\widetilde{D}[t_{1},t_{2}],\quad \widetilde{D^{^{\uparrow }}}[\bullet
,t_{2}]+\widetilde{D^{^{\uparrow }}}[t_{1},\bullet ]).\medskip $ }

{\small \quad $\ \ $If \ ($t$$_{1}$,$t_{2})\in (Ti(M))^{2}$\ \ ($t$$_{1}$\
and $t$$_{2}$\ are inhibited for $M$) }

{\small $\quad \quad \quad \widetilde{D^{\uparrow }}[t_{1},t_{2}]:=MIN(%
\widetilde{D}[t_{1},t_{2}],\quad \widetilde{D^{^{\uparrow }}}[\bullet
,t_{2}]+\widetilde{D^{\uparrow }}[t_{1},\bullet ]).\medskip $ }

{\small $\quad $\quad If $\ (t_{1}\in Ti(M))\wedge (t_{2}\notin Ti(M))$\ \ \
\ (Only $t$$_{1}$\ is inhibited for $M$). }

{\small $\quad \quad \quad \widetilde{D^{\uparrow }}[t_{1},t_{2}]$ $:=MIN(%
\widetilde{D}[t_{1},t_{2}]+\widetilde{D}[t_{f},\bullet ],\quad \widetilde{%
D^{\uparrow }}[\bullet ,t_{2}]+\widetilde{D^{^{\uparrow }}}[t_{1},\bullet
]). $ }

{\small $\quad $\quad If $\ (t_{1}\notin Ti(M))\wedge (t_{2}\in Ti(M))$\ \ \
\ (Only $t$$_{2}$\ is inhibited for $M$) }

{\small {\ $\quad \quad \quad \widetilde{D^{^{\uparrow }}}[t_{1},t_{2}]$ $:=$%
$MIN(\widetilde{D}[t_{1},t_{2}]+\widetilde{\beta }[\bullet ],\quad 
\widetilde{D^{\uparrow }}[\bullet ,t_{2}]+\widetilde{D^{^{\uparrow }}}%
[t_{1},\bullet ]).$ } }
\end{itemize}
\end{itemize}
\end{definition}

If $t$ is an activated transition, then $\widetilde{\beta }[t]$ denotes the
minimal time distance between its firing time and that of any activated
transition. $\widetilde{E}$. Therefore, an activated transition $t_{f}$ is 
\textit{not firable} from $\widetilde{E},$ if $\widetilde{\beta }[t_{f}]<0$.
Further, $\widetilde{\beta }[\bullet ]$ represents the maximal dwelling time
in the class.

It is noteworthy that if $\widetilde{E}$\ is an overapproximation of the
exact\ class $E,$\ then all the transitions firable from $E$\ are also
firable from $\widetilde{E}$. However, a transition which is not firable
from $E$ can, on the other hand, be firable\footnote{%
Conversely, if $t_{f}$\ is not firable from $\widetilde{E},$\ then it is not
firable from\ $E$.} from $\widetilde{E}$. Actually, as the class $\widetilde{%
E}$ contains all the states of $E,$\ we can find at least one state $e$ of $%
\widetilde{E}$ unreachable in $E,$\ such that $e$\ can fire $t_{f}.$

\begin{figure}[h]
\centering%
\begin{tabular}{ll}
\includegraphics[width=7 cm]{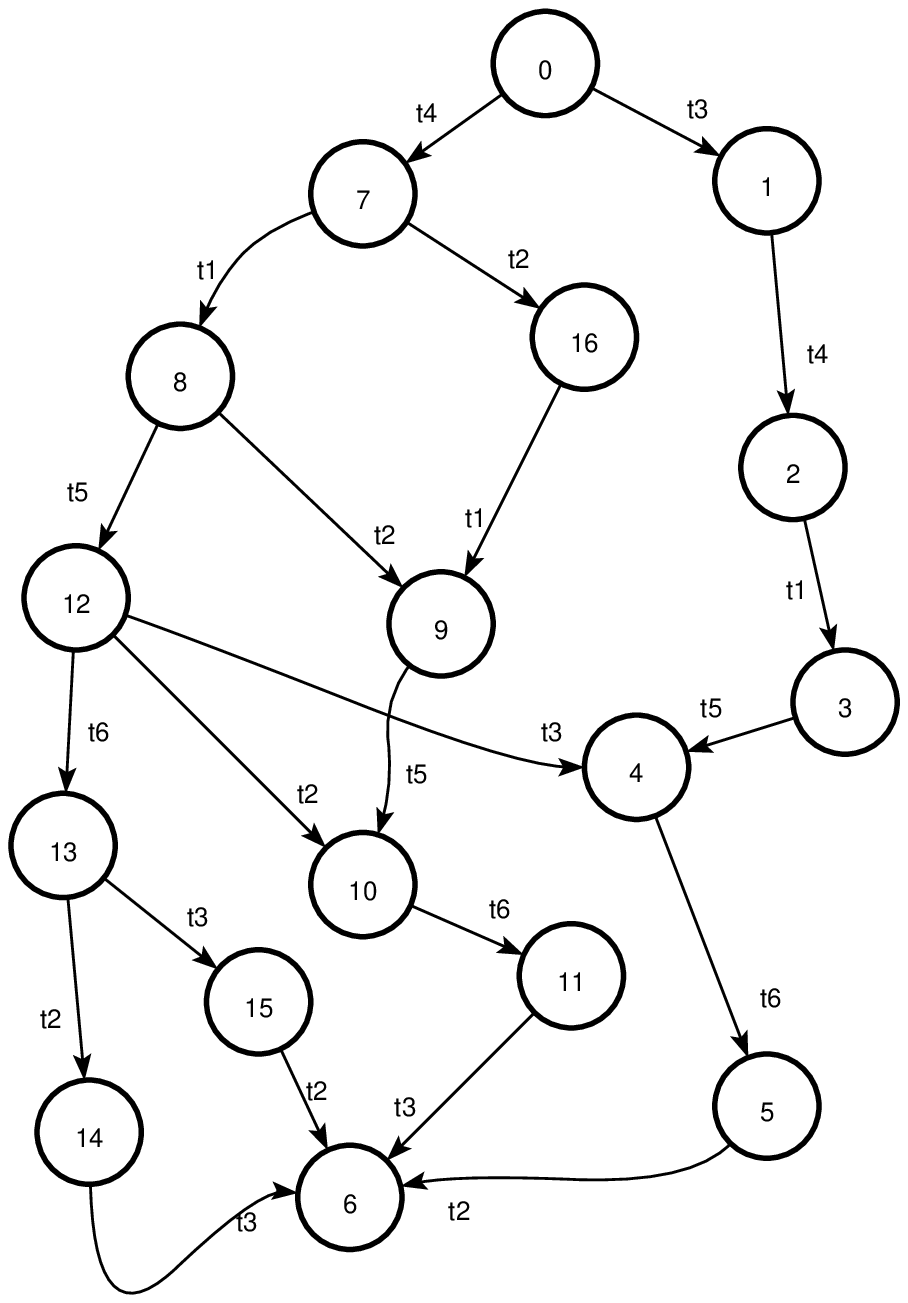} & 
\includegraphics[width=8
cm]{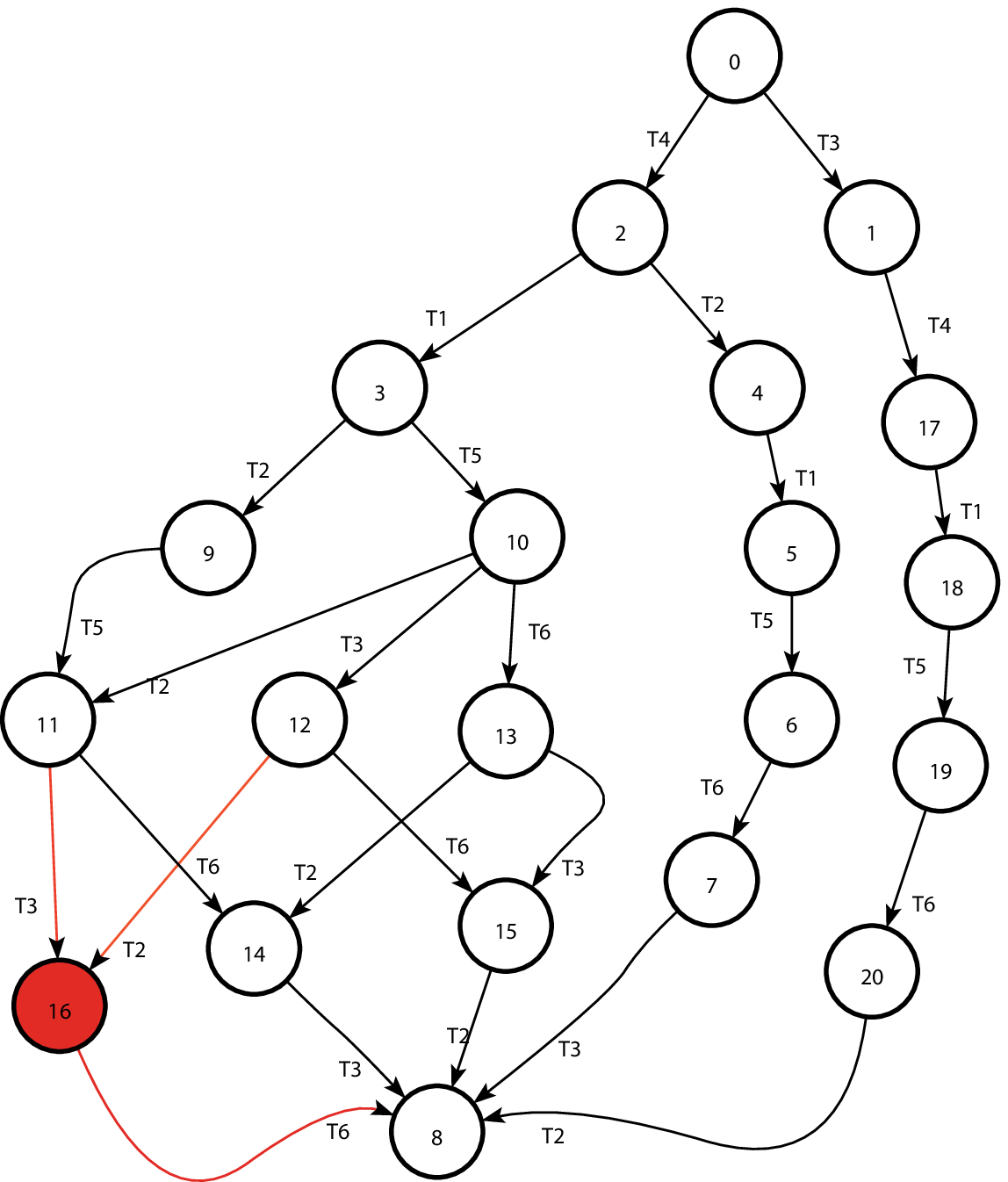} \\ 
\multicolumn{1}{c}{(a)} & \multicolumn{1}{c}{(b)}%
\end{tabular}%
\caption{The exact graph and its DBM overapproximation of the $ITPN$ of
Fig.1 }
\end{figure}

To illustrate both graph constructions, let us consider again the net of 
\textit{Fig 1}. The exact state class graph resulted after applying the
algorithm of Definition 4 is shown in \textit{Fig. 2.a}. Its \textit{DBM}
overapproximation resulted by the application of the algorithm given in
Definition 5 is depicted in \textit{Fig. 2.b}. Hence, the exact graph
contains 17 classes and 22 edges, whereas its \textit{DBM} overapproximated
graph contains 21 classes and 28 edges. By comparing both graphs\footnote{%
The class $\widetilde{E^{n}}$ as well as $E^{n}$ denote the node numbered $n$
in the corresponding graph.}, we notice that the transition $t_{3}$ is
firable from the class $\widetilde{E^{11}}$ in $\widetilde{GR},$ whereas it
is not from $E^{10}$ in $GR$. Moreover, $t_{2}$ is firable from $\widetilde{%
E^{12}}$ whereas it is not from $E^{4}.$ The sequences added in the graph $%
\widetilde{GR}$ due to overapproximation are highlighted in red in \textit{%
Fig 2.b}.

Although the cost of computing the \textit{DBM} overapproximation is low
comparing to the exact construction, it remains that in certain cases the
approximation is too coarse to restore properties of interest and especially
quantitative properties. In actual fact, more transitions remain
persistently enabled along firing sequences more the approximation looses
its precision and therefore includes false behaviors that skew the time
analysis of the system. Besides, these false behaviors may compute an
infinity of unreachable markings while the exact construction is indeed
bounded. Hence, this prevents the \textit{DBM} overapproximation to
terminate while the exact construction may converge.

We investigate in the next section a new approach to compute a tighter 
\textit{DBM} overapproximation. The idea is to restore from previous classes
in the firing sequence time constraints that are used to tighten still more
the \textit{DBM} overapproximation of a class.

\section{Time distance based Approximation of the ITPN State Space}

\begin{tabular}{ccccl}
{\tiny $t_{f}^{1}$} & {\tiny $t_{f}^{2}$} &  & {\tiny $t_{f}^{n}$} & 
{\scriptsize fired transitions} \\ 
{\tiny 0$<$--------$>$} & {\tiny 1 $<$ --------$>$2} & {\tiny $\ .\ .\ .$ }
& {\tiny n-1$<$--------$>$n} & {\scriptsize firing points} \\ 
\multicolumn{1}{l}{{\tiny $M^{0}$}} & \multicolumn{1}{l}{{\tiny $M^{1}\qquad
\ \ \ \ \ \ \ \ M^{2}$}} & {\tiny $\qquad \qquad $} & \multicolumn{1}{l}{%
{\tiny $M^{n-1}$\quad\ \quad\ \quad $M^{n}$\ }} & {\scriptsize reachable
markings} \\ 
\multicolumn{1}{l}{{\tiny $e^{0}$}} & \multicolumn{1}{l}{{\tiny $e^{1}$}%
\qquad\ \ {\tiny $e^{2}$}} & \multicolumn{1}{l}{} & {\tiny $e^{n-1}$\quad\
\quad\ \quad\ \ $e^{n}$} & {\scriptsize reachable states} \\ 
\underline{{\tiny $t_{f}^{1}$}} & \underline{{\tiny $t_{f}^{2}$}} &  & 
\underline{{\tiny $t_{f}^{n}$}} & {\scriptsize firing time distances}%
\end{tabular}%
\bigskip

Let $RT:=(P,T,B,F,M^{0},I,IH)$ be an \textit{Inhibitor arc Time Petri Net. }%
We suppose that a sequence of transitions $S=(t_{f}^{1},..,t_{f}^{n})$ has
been fired in \emph{RT}$.$ The marking and the state reachable at the $%
(j)^{th}$ firing point are denoted $M^{j}$ and $e^{j}$ respectively.%
{\scriptsize {\ } }Therefore, for the firing point $(n)$ we define the
following:

\begin{itemize}
\item The marking reachable at point $(n)$ is denoted by $M^{n}$.

\item The function $Ne^{n}:Te(M^{n})\longrightarrow \left\{ 0,1,..,n\right\}
;~Ne^{n}(t)$ gives, as shown in $Fig.3,$ the number of the firing point that
has enabled the transition $t$ for the last time, provided that $t$ remains
persistently enabled up to the firing point $(n)$. Thereafter, we denote by$%
~[Ne^{n}]$ the set of transition's enabling points reported at the firing
point $(n).$

\item The function $Ni^{n}:Te(M^{n})\longrightarrow \left\{
-1,0,1,..,n\right\} .~Ni^{n}(t)$ gives, as shown in $Fig.3,$ the number of
the firing point that has inhibited the transition $t$ for the last time,
provided that $t$ remains persistently enabled up to the firing point $(n)$.
We have $Ni^{n}(t)=-1$ if $t$ has never been inhibited since its last
enabling point. Thereafter, we denote by$~[Ni^{n}]$ the set of transition's
inhibiting points reported at the firing point $(n).$

\item The function $Na^{n}:Te(M^{n})\longrightarrow \left\{
-1,0,1,..,n\right\} .~Na^{n}(t)$ gives, as shown in $Fig.3,$ the number of
the firing point that has activated the transition $t$ for the last time,
provided that $t$ remains persistently enabled up to the firing point $(n)$.
We have $Na^{n}(t)=-1$ if $t$ has never been activated since its last
enabling point. Thereafter, we denote by$~[Na^{n}]$ the set of transition's
activating points reported at the firing point $(n).$

\item We denote thereafter by $Point^{n}$ the set $[Ne^{n}]\cup \lbrack
Ni^{n}]\cup \lbrack Na^{n}]-\{-1\}.$
\end{itemize}

\begin{figure}[tp]
\centering\includegraphics[width=12 cm]{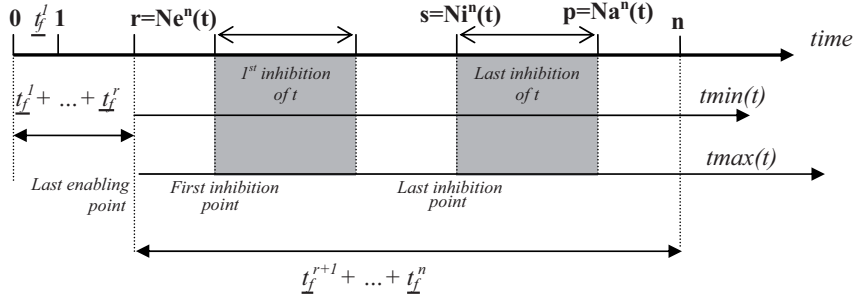}
\caption{Last enabling inhibiting and activating points of a transition.}
\end{figure}

Let us consider the firing of a sequence of transitions $S=\left(
t_{f}^{1},..,t_{f}^{n}\right) $ in the graph $GR$. The sequence $S$
describes a path in the graph $GR$ going from the node representing the
class $E^{0}$ to the node which represents the class $E^{n}$. We introduce
next the time distance system that encodes the quantitative properties of
some subsequences of $S$.

\begin{definition}
{\small {\ \textbf{\ }\ Let $E^{n}=(M^{n},D^{n})$\ be a class reachable in $%
GR,$ after firing the sequence\ $S=\left( t_{f}^{_{1}},..,t_{f}^{n}\right) .$
For point $(n),$ we define the time distance system, noted $DS^{n},$ as
follows:\newline
$DS^{n}=\left\{ 
\begin{tabular}{l}
{\Huge $\wedge $}$_{\forall i\in Point^{n}},-DS^{n}[n,i]\leq \underline{%
t_{f}^{_{i+1}}}+..+\underline{t_{f}^{n}}\leq DS^{n}[i,n]$\newline
\\ 
{\Huge $\wedge $}$_{\forall i\in Point^{n}\cup \{n\}}${\Huge $\wedge $}$%
_{\forall t\in Te(M^{n})},-DS^{n}[t,i]\leq \underline{t_{f}^{_{i+1}}}+..+%
\underline{t_{f}^{n}}+\underline{t}\leq DS^{n}[i,t]$%
\end{tabular}%
\right. $\ } }
\end{definition}

More concretely, if $t$ is an enabled transition for $E^{n}$, then $%
DS^{n}[t,i]$ represents the opposite value of the minimum residual time of $%
t $ computed from the firing point $(i),$ whereas $DS^{n}[i,t]$ denotes its
maximum residual time relatively to the firing point $(i)$. Moreover, $%
DS^{n}[i,n]$ (respectively, $DS^{n}[n,i]),$ denotes the maximum time
distance (respectively, the opposite value of the minimum time distance),
between the firing points $(i)$ and $(n)$. The coefficients of the system $%
DS^{0}$ are defined as follows:\ We have $Point^{n}=\{0\};$\newline
$\forall t\in Te(M^{0}),$ \qquad $DS^{0}[0,0]=0,$ $DS^{0}[0,t]=tmax(t),$ $%
DS^{0}[t,0]=-tmin(t).$ $\ \ \ $

Thereafter, we encode the system $DS^{n}$ as four matrices. For instance,
the coefficients of the system $DS^{0}$ of the $ITPN$ of \textit{Fig.1} are
given in \textit{Tab.2}.

\begin{table}[h]
\caption{ The time distance system at firing point $(0)$}\centering%
\begin{tabular}{l}
\begin{tabular}{|c|c|c|c|}
\hline
$DS^{0}[i,t]$ & $t_{1}$ & $t_{3}$ & $t_{4}$ \\ \hline
$0$ & 3 & 4 & 2 \\ \hline
\end{tabular}%
\medskip\ 
\begin{tabular}{|c|c|c|c|}
\hline
$DS^{0}[t,i]$ & $t_{1}$ & $t_{3}$ & $t_{4}$ \\ \hline
$0$ & -3 & -2 & 0 \\ \hline
\end{tabular}%
\medskip\ 
\begin{tabular}{|c|c|}
\hline
$DS^{0}[i,n]$ & $0$ \\ \hline
$0$ & 0 \\ \hline
\end{tabular}
\begin{tabular}{|c|c|}
\hline
$DS^{0}[n,i]$ & $0$ \\ \hline
$0$ & 0 \\ \hline
\end{tabular}%
\end{tabular}%
\end{table}

Next definition shows how the system $DS^{n}$ can be determined recursively
as a result of solving a general polyhedral system.

\begin{definition}
{\small {\ {\ \textit{Let }$E^{n-1}=(M^{n-1},D^{n-1})$\textit{\ be a class
reachable in }$GR$\textit{\ and let }$\mathit{DS}^{n-1}$\textit{\ be the
time distance system associated with the class }$E^{n-1}$\textit{. Let us
consider }$E^{n}=(M^{n},D^{n})$\textit{\ be the class reachable from }$%
E^{n-1}$\textit{\ after firing the transition }$t_{f}^{n}$\textit{. The time
distance system }$DS^{n}$\textit{\ associated with }$E^{n}$\textit{\ can be
worked out recursively from the systems }$DS^{n-1},D^{n-1}$\textit{\ \ as
follows: } } }}

\begin{enumerate}
\item {\small Compute the function $Ne$ $^{n}$\ as follows: $\forall t\in
Te(M^{n})$}\newline
{\small \ If $t\in New(M^{n})$ then $Ne^{n}(t):=n$ \qquad else $%
Ne^{n}(t):=Ne^{n-1}(t).$ }\newline
{\small Compute the function $Ni$$^{n}$\ as follows: }\newline
{\small If $Ne^{n}(t)=n$ then if $t\in Ti(M$$^{n})$ then $Ni^{n}(t):=n$ else 
$Ni^{n}(t):=-1.$\qquad }

\quad {\small else if $t\in Ta(M$ $^{n-1})$ $\wedge t\in Ti(M$$^{n})$ then $%
Ni^{n}(t):=n.$ }

\quad \quad \quad \quad \quad \quad {\small else $Ni^{n}(t):=Ni^{n-1}(t)$. }%
\newline
{\small Compute the function $Na$$^{n}$\ as follows:}\newline
{\small If $Ne^{n}(t)=n$ then if $t\in Ta(M$$^{n})$ then $Na^{n}(t):=n$ else 
$Na^{n}(t):=-1.$\qquad }

\quad {\small else if $t\in Ta(M$$^{n})$$\wedge t\in Ti(M$$^{n-1})$ then $%
Na^{n}(t):=n.$ }

\quad \quad \quad \quad \quad \quad {\small else $Na^{n}(t):=Na^{n-1}(t)$. }%
\newline

\item {\small Augment the system $D^{n-1}$\ with the firing constraints of $%
t_{f}^{n}$\ that we write $D_{a}^{n-1}=$ $D_{a}^{n-1}\wedge (\forall t\in
Ta(M^{n-1}),\quad \underline{t_{^{f}}^{n}}\leq \underline{t})$ }

\item {\small In the system $D_{a}^{n-1}\wedge DS^{n-1}$\ rename each
variable $\underline{t}$\ related to\ an activated transition which is
persistent for M$^{n}$\ with $\underline{t}^{\prime }+\underline{t_{f}^{n}}$%
. For inhibited transitions, rename the related variable $\underline{t}$\
with $\underline{t}^{\prime }$. }

\item {\small In the resulted system and by intersection of the constraints,
remove the variables related to disabled transitions\ and determine the
constraints of $DS^{n}$. \ \medskip }

\item {\small In the obtained system, add constraints related to newly
enabled transitions, as follows : $\forall t\in New(M^{n}),$ $\forall i\in
Point^{n}\cup \{n\}$\newline
$-DS^{n}[n,i]-tmin(t)\leq \underline{t_{f}^{_{i+1}}}+..+\underline{t_{f}^{n}}%
+\underline{t}\leq DS^{n}[i,n]+tmax(t).$ }
\end{enumerate}
\end{definition}

The computation of the system $DS^{n}$ is very complex as it needs at each
step to manipulate a global system which may contain polyhedral constraints.
Concretely, if the latter appears in $D^{n-1}$ then the cost of computing $%
DS^{n}$\ is exponential on the number of variables, otherwise it is
polynomial. However, in most of the cases, polyhedral constraints do not
affect the computation of the time distances. Therefore, to alleviate the
computation effort, the idea is to leave out systematically such constraints
during the process (keeping only the $DBM$ system to represent the space of
the class $E^{n-1}),$ with a risk however to compute in certain cases an
overapproximation of the system $DS^{n}.$ The resulted system obtained by 
\textit{DBM} restriction is noted thereafter $\widetilde{DS^{n}}$ and we
have $\widetilde{DS^{0}}=DS^{0}$. The next proposition provides an algorithm
to compute recursively and efficiently the coefficients of the system $%
\widetilde{DS_{n}}$ in the context of the DBM overapproximated graph that we
aim to compute, noted $\widetilde{GRc}$. However the same algorithm can be
applied in the context of the exact graph $GR$ while restricting the system $%
D$ to $\overrightarrow{D}$ ( $\overrightarrow{D}$ is the tightest DBM
overapproximation that one can compute from $D$).

\begin{proposition}
{\small {\textit{Let the graph }}}$\widetilde{{\small {GRc}}}$ {\small be a
DBM overapproximation of the graph }${\small GR}${\small . Let{\textit{\ }}}$%
\widetilde{{\small {E}}_{{\small {c}}}^{{\small {n-1}}}}${\small {$=(M^{n-1},%
\widetilde{D_{c}^{n-1}})$\textit{\ be a class reachable in }}}$\widetilde{%
{\small {GRc}}}${\small {$,$\textit{\ from the initial class\ after firing
the sequence }$S=\left( t_{f}^{_{1}},..,t_{f}^{n-1}\right) $}.{\ \textit{Let 
}$\widetilde{\mathit{DS}^{n-1}}$\textit{\ be the DBM overapproximated time
distance system associated with the class }}}$\widetilde{{\small {E}}_{%
{\small {c}}}^{{\small {n-1}}}}${\small {\textit{. Let us consider }}}$%
\widetilde{{\small {E}}_{{\small {c}}}^{{\small {n}}}}${\small {$=(M^{n},%
\widetilde{{D}_{{c}}^{{n}}})$\textit{\ the class reachable from }}}$%
\widetilde{{\small {E}}_{{\small {c}}}^{{\small {n-1}}}}${\small {\textit{\
after firing the transition }$t_{f}^{n}$\textit{. The DBM overapproximated
time distance system }}}$\widetilde{{\small {DS^{n}}}}${\small {\textit{\
associated with }}}$\widetilde{{\small {E}}_{{\small {c}}}^{{\small {n}}}}$%
{\small {\textit{\ can be computed recursively from previous systems in the
sequence }$\mathit{S}$\textit{,\ as follows:} }} {\scriptsize {\ }}

\begin{itemize}
\item {\scriptsize {\small Compute the function $Ne^{n},Ni^{n}$ and $Na^{n}$%
\ as in Definition.7. } }

\item {\scriptsize {\small The coefficients of the system $\widetilde{DS^{n}}
$\ are computed by using the following formulae: } }

{\scriptsize {\small $\forall i\in Point$}}$^{{\scriptsize {\small n}}}$%
{\scriptsize {\small \ } }

{\scriptsize {\small $\quad \widetilde{DS^{n}}[i,n]:=\lambda ^{n-1}[i];$ } }

{\scriptsize {\small $\quad \widetilde{DS^{n}}[n,i]:=\widetilde{DS^{n-1}}%
[t_{f}^{n},i];$ } }

{\scriptsize {\small $\quad \widetilde{DS^{n}}[n,n]:=0;$ } }

{\scriptsize {\small such that }$\forall i\in Point^{n}\cup \{n\},$ $\lambda
^{n}[i]=\underset{t\in Ta(M^{n})}{MIN}\left\{ \widetilde{DS^{n}}%
[i,t]\right\} ${\small \ } }

{\scriptsize {\small $\forall t\in Te(M^{n}),$\ $\forall i\in Point$}$^{%
{\small n}}$ }

{\scriptsize {\small \ \ If $Ne^{n}(t)=n$\ ($t$\ is newly enabled) } }

{\scriptsize {\small $\quad \quad \quad 
\begin{tabular}{ll}
$\widetilde{DS^{n}}[i,t]:=\widetilde{DS^{n}}[i,n]+tmax(t);$ & $\quad 
\widetilde{DS^{n}}[t,i]:=\widetilde{DS^{n}}[n,i]-tmin(t);$ \\ 
$\widetilde{DS^{n}}[n,t]:=tmax(t);$ & $\quad \widetilde{DS^{n}}%
[t,n]:=-tmin(t);$%
\end{tabular}%
$} {\small $\quad \quad $ } {\small $\quad \quad \quad $ }\newline
}

{\scriptsize {\small \ \ If $Ne^{n}(t)\neq n$\ ($t$\ is persistent) } }

{\scriptsize {\small \quad\ \ \ If $t\notin Ti(M^{n-1})$\ $\ (t$ is not
inhibited for $M$$^{n-1}$)}$,${\small \ } }

{\scriptsize \ {\small \quad\ \ \ Let }}${\scriptsize {\small r=Ne}}^{%
{\scriptsize {\small n}}}{\scriptsize {\small (t),s=Ni}}^{{\scriptsize 
{\small n-1}}}{\scriptsize {\small (t)}}$ {\small and}{\scriptsize \ }$%
{\scriptsize {\small p=Na}}^{{\scriptsize {\small n}}}{\scriptsize {\small %
(t).}}${\scriptsize \ }

{\scriptsize {\small $\quad \quad \quad \widetilde{DS^{n}}[i,t]:=MIN\left( 
\begin{tabular}{l}
$\left\{ 
\begin{tabular}{ll}
$\widetilde{DS^{s}}[i,t]+\lambda ^{n-1}[s]+\widetilde{DS^{n}}[n,p]$ & if $%
0\leq i\leq s\leq p$ \\ 
$\widetilde{DS^{i}}[i,t]+\widetilde{DS^{n}}[i,n]+\widetilde{DS^{n}}[n,p]$ & 
if $0\leq s\leq i\leq p$%
\end{tabular}%
\right. $ \\ 
$\widetilde{DS^{n-1}}[i,t]$ \\ 
$\widetilde{DS^{n}}[i,n]+\widetilde{DS^{n-1}}[n-1,t]+\widetilde{DS^{n-1}}%
[t_{f}^{n},n-1]$%
\end{tabular}%
\right. $}\newline
}

{\scriptsize {\small $\quad \quad \quad \widetilde{DS^{n}}[t,i]:=MIN\left( 
\begin{tabular}{l}
$\left\{ 
\begin{tabular}{ll}
$\widetilde{DS^{s}}[t,i]+\widetilde{DS^{n-1}}[t_{f}^{n},s]+\widetilde{DS^{n}}%
[p,n]$ & if $0\leq i\leq s\leq p$ \\ 
$\widetilde{DS^{i}}[t,i]+\widetilde{DS^{n}}[n,i]+\widetilde{DS^{n}}[p,n]$ & 
if $0\leq s\leq i\leq p$%
\end{tabular}%
\right. $ \\ 
$\widetilde{DS^{n-1}}[t,i]$$\quad $ \\ 
$\widetilde{DS^{n}}[n,i]+MIN(0,$ $\widetilde{DS^{n-1}}[t,n-1]+\lambda
^{n-1}[n-1])$%
\end{tabular}%
\right. $} }

{\scriptsize {\small $\quad \quad \quad \widetilde{DS^{n}}[t,n]:=\quad
MIN\left\{ 
\begin{tabular}{l}
$\widetilde{\beta _{c}^{n-1}}[t]$ \\ 
$MIN(0,$ $\widetilde{DS^{n}}[t,r]+\widetilde{DS^{n}}[r,n])$%
\end{tabular}%
\right. ;$ } }

{\scriptsize {\small $\quad \quad \quad \widetilde{DS^{n}}[n,t]:=\quad
MIN\left\{ 
\begin{tabular}{l}
$\widetilde{D_{c}^{n-1}}[t_{f}^{n},t]$ \\ 
$\widetilde{DS^{n}}[r,t]+\widetilde{DS^{n}}[n,r]$%
\end{tabular}%
\right. ;\smallskip $ } }

{\scriptsize {\small $\quad \quad $If $t\in Ti(M^{n-1})$\ $\ (t$ is
inhibited for $M$$^{n-1}$), } }

{\scriptsize {\small $\quad \quad \quad \quad $Let $s=Ni^{n}(t)$ and $%
r=Ne^{n}(t).$} }

{\scriptsize {\small $\quad \quad $$\quad \widetilde{DS^{n}}[i,t]:=$}}

{\scriptsize {\small $\quad \quad \quad \quad MIN$$\left( 
\begin{tabular}{l}
$\left\{ 
\begin{tabular}{ll}
$\widetilde{DS^{s}}[i,t]+\widetilde{DS^{n}}[s,n]$ & if $i\leq s$ \\ 
$\widetilde{DS^{i}}[i,t]+\widetilde{DS^{n}}[i,n]$ & if $s\leq i$%
\end{tabular}%
\right. \smallskip $ \\ 
$\widetilde{DS^{n-1}}[i,t]+\lambda ^{n-1}[n-1]$%
\end{tabular}%
\right. $ } }

{\scriptsize {\small $\quad $$\quad \quad \widetilde{DS^{n}}[t,i]:=$}}

{\scriptsize {\small $\quad \quad \quad \quad MIN$$\left( 
\begin{tabular}{l}
$\left\{ 
\begin{tabular}{ll}
$\widetilde{DS^{s}}[t,i]+\widetilde{DS^{n}}[n,s]$ & if $i\leq s$ \\ 
$\widetilde{DS^{i}}[t,i]+\widetilde{DS^{n}}[n,i]$ & if $s\leq i$%
\end{tabular}%
\right. \smallskip $ \\ 
$\widetilde{DS^{n-1}}[t,i]+\widetilde{DS^{n-1}}[t_{f}^{n},n-1]$%
\end{tabular}%
\right. $ \smallskip } }

{\scriptsize {\small $\quad \quad $$\quad \widetilde{DS^{n}}[n,t]:=\quad
MIN\left( 
\begin{tabular}{l}
$\widetilde{DS^{n}}[r,t]+$$\widetilde{DS^{n}}[n,r]$ \\ 
$\widetilde{DS^{n-1}}[n-1,t]$ \\ 
$\widetilde{D^{n-1}}[t_{f}^{n},t]+$$\lambda ^{n-1}[n-1];$%
\end{tabular}%
\right. $ } }

{\scriptsize {\small $\quad \quad \quad $$\widetilde{DS^{n}}[t,n]:=\quad
MIN\left( 
\begin{tabular}{l}
$MIN(0,\widetilde{DS^{n}}[t,r]+\widetilde{DS^{n}}[r,n])$ \\ 
$\widetilde{DS^{n-1}}[t,n-1]$ \\ 
$\widetilde{DS^{n-1}}[t_{f}^{n},n-1]+\widetilde{\beta _{c}^{n-1}}[t]$%
\end{tabular}%
\right. \medskip $ } }

{\small such that} {\small $\forall t\in Te(M^{n-1}),\quad \widetilde{\beta
_{c}^{n-1}}[t]=\underset{\forall t^{\prime }\in Ta(M^{n-1})}{MIN}\left\{ 
\widetilde{D_{c}^{n-1}}[t,t^{\prime }]\right\} .$}
\end{itemize}
\end{proposition}

The previous proposition provides an efficient algorithm to compute an
overapproximation of the system $DS^{n}.$ For this effect, the algorithm
starts to determine the set $Point^{n}${\small $.$} Then it calculates the
coefficients $\widetilde{DS^{n}}[i,n]$ and $\widetilde{DS^{n}}[n,i]$ for
each point $i\in Point^{{\small n}}${\small $.$} Then for each enabled
transition $t$, the algorithm computes the other coefficients following the
cases:

\begin{figure}[b]
\centering%
\begin{tabular}{c}
\includegraphics[width=12 cm]{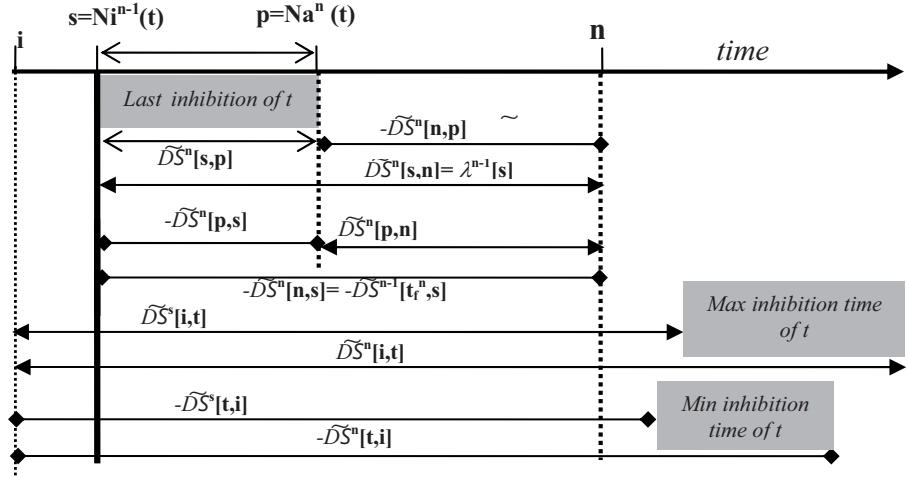} \\ 
(a) \\ 
\includegraphics[width=12.5 cm]{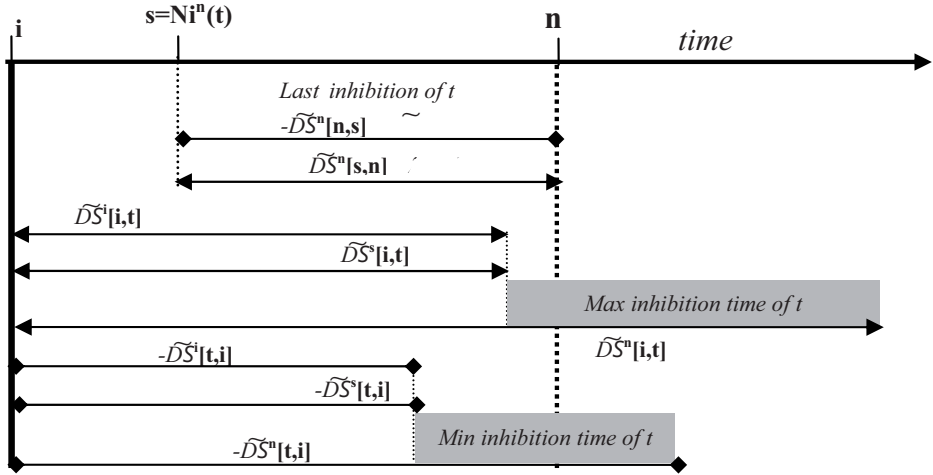} \\ 
(b)%
\end{tabular}%
\caption{Time distance Computation.}
\end{figure}

\begin{itemize}
\item When dealing with newly enabled transitions, the formulae are obvious
and are the same for inhibited and activated transitions.

\item When handling persistent transitions, the algorithm proceeds first to
compute the distances $\widetilde{DS^{n}}[i,t]$ and $\widetilde{DS^{n}}[t,i]$
for each point $i\in Point^{{\small n}}$. It is noteworthy that the previous
distances are more likely to maintain their values along a firing sequence
as long as $t$ is not inhibited in the sequence. However, if $t$ becomes
inhibited then these distances increase by the time elapsed during its
inhibition. Therefore, if a transition is activated for the point $(n-1)$,
and $t$ has never been inhibited since its last enabling point ($s=-1$),
then both distances are more likely to maintain their old values, even
decreasing in very rare cases if there is state space restriction (see the
last two items of the \textit{MIN}). However, if the transition $t$ has been
inhibited at least once since its last enabling point ($s\geq 0$), then the
duration of its last inhibition time should be re-calculated at each new
reachable point to better approximate these distances. In actual fact,
because of space restriction the inhibition times of $t$ may decrease even
after that $t$ has been activated. As a result, the interval $[-\widetilde{%
DS^{n}}[t,i],\widetilde{DS^{n}}[i,t]]$ may only narrow along a firing
sequence as long as $t$ remains activated. For this purpose, we need to
restore some time information computed earlier in the sequence at points ($s$%
) and ($i$). For instance, if the point $(i)$ occurs first in the firing
sequence, then the distance $\widetilde{DS^{n}}[i,t]$ is likely to be equal
to the same distance computed at point $s$, $\widetilde{DS^{s}}[i,t]$
augmented with the maximal inhibition time of $t$, namely\footnote{%
Note that we use rather $\lambda ^{n-1}[s]$ than $\widetilde{DS^{n}}[s,n]$
in the formula, because the point $(s)$ may be not defined in $Point^{n}$ if 
$t$ is inhibited at point $(n)$.} $\lambda ^{n-1}[s]+\widetilde{DS^{n}}[n,p]$
(see \textit{Fig 3.a},). Otherwise, if the point $(i)$ occurs during the
inhibition time of $t$, then the distance $\widetilde{DS^{n}}[i,t]$ is
likely to be equal to the same distance computed at point $(i),$ $\widetilde{%
DS^{i}}[i,t]$ augmented with the maximal inhibition time of $t$ from point $%
(i)$ to $(p):\widetilde{DS^{n}}[i,n]+\widetilde{DS^{n}}[n,p]$.

In case that $t$ is inhibited for the point $(n-1)$, we follow the same
approach as previously to compute the same distances. However, in this case
the adjustment of the approximation is carried out during the inhibition
time of the transition $t$. At each new firing point, the residual time of
an inhibited transition should increase with the dwelling time measured at
point $(n-1)$. Furthermore, as shown in \textit{Fig 3.b}, if $i$ occurs
before $s$, then this distance should not surpass the residual time of the
transition reported at point $(s)$ augmented with the inhibition time
elapsed from $(s)$ till the current point $(n)$.

\item The algorithm ends the process by calculating the coefficients $%
\widetilde{DS^{n}}[n,t]$ and $\widetilde{DS^{n}}[t,n]$. As these
coefficients denote the same distances as respectively $\widetilde{D_{c}^{n}}%
[\bullet ,t]$ and $\widetilde{D_{c}^{n}}[\bullet ,t]$ already defined in a
classical $DBM$ system. Therefore, their computation is worked out also by
using the formulae of Definition 5, already established in \cite{abdelli}.
Better still, new formulae are added to tighten still more their
approximation.
\end{itemize}

We propose thereafter to exploit the time distance system in the computation
of an overapproximation of the state class graph of an$\ ITPN$. \ The
proposition.1 shows that by overapproximating the computation of the system $%
DS^{n},$ we reduce the effort of its computation to a polynomial time. From
this system, we are able to restore some time information that makes it
possible to compute a $DBM$ overapproximation that is tighter than that of
other approaches \cite{abdelli}\cite{Bucci}\cite{IHTPN}. Formally, the time
distance based approximation of the graph $GR$ is built as follows:

\begin{definition}
{\small {\ The time distance based approximation graph of an $ITPN$, denoted
by $\widetilde{GRc}$ is the tuple $(\widetilde{CEc},\widetilde{E_{c}^{0}}%
,\leadsto )$\ such that: }}

\begin{itemize}
\item {\small {\ $\widetilde{CEc}$\ is the set of approximated classes
reachable in $\widetilde{GRc\text{ }};$\ } }

\item {\small {$\widetilde{E_{c}^{0}}=(M^{0},Ne^{0},Ni^{0},Na^{0},\widetilde{%
DS^{0}},\widetilde{D_{c}^{0}})\in \widetilde{CEc}$ is the initial class such
that }$\widetilde{{DS}}${$^{0}=DS^{0}$ and }$\widetilde{{D_{c}^{0}}}${\ is
the system }$\forall t,t^{\prime }\in Te(M^{0}),\underline{t}^{\prime }-%
\underline{t}\leq tmax(t^{\prime })-tmin(t)$ }

\item {\small $\leadsto ${\ is a transition relation between approximated
classes defined on $\widetilde{CEc}\times T\times \widetilde{CEc},$\ such
that $((M^{n-1},Ne^{n-1},Ni^{n-1},Na^{n-1},\widetilde{DS^{n-1}},\widetilde{%
D_{c}^{n-1}}),t_{f}^{n},(M^{n},Ne^{n},Ni^{n},Na^{n},\widetilde{DS^{n}},%
\widetilde{D_{c}^{n}}))$ $\ \in \leadsto ,$ iff: } }

\begin{description}
\item[(i)] {\small $t_{f}^{n}$$\in Ta(M^{n-1}).$ }

\item[(ii)] {\small $\widetilde{\beta _{c}^{n-1}}[t_{f}^{n}]\geq 0$.}
\end{description}

{\small The new class is computed as follows: }

\begin{itemize}
\item {\small $\forall p\in P,$ $%
M^{n}(p):=M^{n-1}(p)-B(p,t_{f}^{n})+F(p,t_{f}^{n})\smallskip .$ }

\item {\small Compute the function $Ne^{n},Ni$}$^{{\small n}}${\small \ and $%
Na^{n}$\ as in Definition.7. }

\item {\small Compute the coefficients of the system $\widetilde{DS^{n}}$\
as in Proposition 1: }

\item {\small The DBM system $\widetilde{D_{c}^{n}}$\ is obtained as follows
: }

{\small $\forall (t,t^{\prime })\in (Te(M^{n-1}))^{2}\wedge (t\neq t^{\prime
})$ }

{\small \quad If $t$ or $t^{\prime }$ are newly enabled. }

{\small $\ \ \quad \widetilde{D_{c}^{n}}[t,t^{\prime }]:=\widetilde{DS^{n}}%
[n,t^{\prime }]+\widetilde{DS^{n}}[t,n]$.\medskip \medskip }

{\small \quad If $t$\ and $t^{\prime }$\ are persistent. }

{\small $\quad $\quad If ($t,t^{\prime })\notin (Ti(M^{n-1}))^{2}$\ \ ($t$\
and $t^{\prime }$\ are not inhibited for $M^{n-1}$) }

{\small $\quad \quad \quad \widetilde{D_{c}^{n}}[t,t^{\prime }]:=MIN(%
\widetilde{D_{c}^{n-1}}[t,t^{\prime }],\quad \alpha ^{n}[t,t^{\prime
}]).\medskip $\ }

{\small \quad $\ \ $If \ ($t$,$t^{\prime })\in (Ti(M^{n-1}))^{2}$\ \ ($t$\
and $t^{\prime }$\ are inhibited for $M^{n-1}$) }

{\small $\quad \quad \quad \widetilde{D_{c}^{n}}[t,t^{\prime }]:=MIN(%
\widetilde{D_{c}^{n-1}}[t,t^{\prime }],\quad \alpha ^{n}[t,t^{\prime
}]).\medskip $\ }

{\small $\quad $\quad If $\ (t\in Ti(M^{n-1}))\wedge (t^{\prime }\notin
Ti(M^{n-1}))$\ \ \ \ (Only $t$\ is inhibited for $M^{n-1}$). }

{\small $\quad \quad \quad \widetilde{D_{c}^{n}}[t,t^{\prime }]$\ $:=MIN(%
\widetilde{D_{c}^{n}}[t,t^{\prime }]+\widetilde{D_{c}^{n-1}}%
[t_{f}^{n},\bullet ],\quad \alpha ^{n}[t,t^{\prime }]).$\ }

{\small $\quad $\quad If $\ (t\notin Ti(M^{n-1}))\wedge (t^{\prime }\in
Ti(M^{n-1}))$\ \ \ \ (Only $t^{\prime }$\ is inhibited for $M^{n-1}$) }

{\small \ $\quad \quad \quad \widetilde{D_{c}^{n}}[t,t^{\prime }]$\ $:=$$MIN(%
\widetilde{D_{c}^{n-1}}[t,t^{\prime }]+\lambda ^{n-1}[n-1],\quad \alpha
^{n}[t,t^{\prime }]).$\medskip }

{\small such that $\alpha ^{n}[t,t^{\prime }]$\ =$\underset{i\in
Point^{n}\cup \{n\}}{MIN}\left( \widetilde{DS^{n}}[i,t^{\prime }]+\widetilde{%
DS^{n}}[t,i]\right) $ }
\end{itemize}
\end{itemize}
\end{definition}

The previous definition provides an algorithm to compute a \textit{DBM}
overapproximation of an \textit{ITPN}. To avoid redundancy, each computed 
\textit{DBM} system of a reachable class, noted $\widetilde{D_{c}^{n}},$ is
reduced to the constraints $t^{\prime }-t\leq \widetilde{D_{c}^{n}}%
[t,t^{\prime }].$ Note that the other constraints of type $-\widetilde{%
D_{c}^{n}}[t,\bullet ]\leq t\leq \widetilde{D_{c}^{n}}[\bullet ,t]$ are
already computed in the system $\widetilde{DS^{n}}$ as $-\widetilde{DS^{n}}%
[t,n]\leq t\leq \widetilde{DS^{n}}[n,t].$ Comparatively to the construction
of the graph $\widetilde{GR}$ given in Definition.5, the class is extended
to the parameters $Ni^{n}$, $Na^{n}$, $Ni^{n}$ and $\widetilde{DS^{n}}$. The 
\textit{DBM} system $\widetilde{D_{c}^{n}}$ computed thereof is used in the
firing and class' equivalence tests. It is noteworthy that the same firing
condition is used in both constructions. However, the computation of the
coefficients of $\widetilde{D_{c}^{n}}$ are better approximated than those
of the system $\widetilde{D^{n}}$. First of all, as it is given in
Proposition 1, the maximal and the minimal residual times of an enabled
transition use formulae that are more precise than those provided in
Definition 5. As a result, the \textit{DBM} coefficients $\widetilde{%
D_{c}^{n}}[t,t^{\prime }]$ are more precise too. This makes it possible to
tighten still more the approximation and therefore to avoid the generation
of additional sequences that stand in the graph $\widetilde{GR}$. The
resulted graph $\widetilde{GRc}$\ is therefore more precise than $\widetilde{%
GR}$. However, the cost of computing the system $\widetilde{D_{c}^{n}}$ is
slightly higher than $\widetilde{D^{n}}$ as it requires also to consider the
computation effort of the system $\widetilde{DS^{n}}$. Nevertheless, the
total cost of computing a class in $\widetilde{GRc}$ remains polynomial and
equal to $o(m^{2}l+m^{2}+ml),$ where $m$ and $l$ denote respectively the
number of enabled transitions\ and the number of reported points. We need to
prove now formally that the construction of the $\widetilde{GRc}$ computes
in all cases an overapproximation of the exact graph $GR,$ which remains
always tighter than the graph $\widetilde{GR}$.

\begin{theorem}
Let $RT$ be an ITPN and $\widetilde{GR}=(\widetilde{CE},(M^{0},\widetilde{%
D^{0}}),\rightsquigarrow ),$ $\widetilde{GRc}=(\widetilde{CEc}%
,(M^{0},Ne^{0},Ni^{0},Na^{0},\widetilde{DS^{0}},\widetilde{D_{c}^{0}}%
),\leadsto )$\ and $GR=(CE,(M^{0},D^{0}),\longmapsto )$ the graphs build on $%
RT$: $\widetilde{GR}$ is an overapproximation of the graph $\widetilde{GRc}$
and the latter is an overapproximation of the exact graph $GR$.
\end{theorem}

\begin{pf}
{\small The proof is given in Appendix.}
\end{pf}

The previous theorem establishes that the algorithm of $Definition.8$
computes a more precise graph than that computed by using other $DBM$
overapproximation approaches \cite{abdelli}\cite{IHTPN}\cite{Bucci}. As a
result, the size of the graph is reduced since additional sequences might be
fired when using classical DBM approximations whereas they are not in $%
\widetilde{GRc}$ as well as in $GR$. To advocate the benefits of the defined
construction, let us consider again the $ITPN$\ of \textit{Fig 1}. As shown
in \textit{Fig.5}, the obtained graph $\widetilde{GRc}$ is much compact than 
$\widetilde{GR}$ and contains $19$ classes and $25$ edges. Moreover, some of
the additional sequences reported in $\widetilde{GR}$ due to
overapproximation are removed in $\widetilde{GRc}.$

\begin{figure}[h]
\centering\includegraphics[width=9 cm]{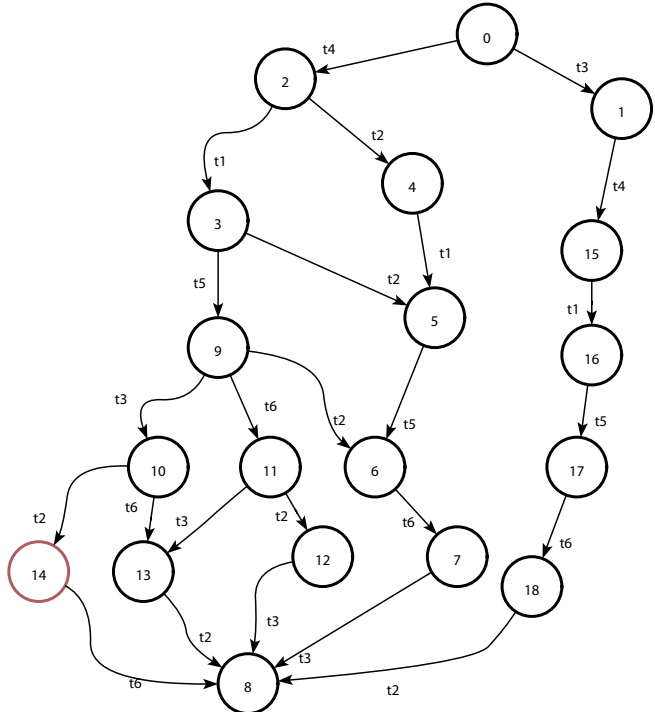}
\caption{Time Distance based approximation graph.}
\end{figure}

For example, let us consider the firing sequence $\widetilde{E_{c}^{0}}%
\overset{t_{4}}{\leadsto }\widetilde{E_{c}^{2}}\overset{t_{1}}{\leadsto }$ $%
\widetilde{E_{c}^{3}}\overset{t_{2}}{\leadsto }$ $\widetilde{E_{c}^{5}}%
\overset{t_{5}}{\leadsto }\widetilde{E_{c}^{6}}$ already discussed in page
10. After firing the transition $t_{4}$ from the initial class, we reach the
class $\widetilde{E_{c}^{2}}$ where $t_{3}$ is inhibited for the first time.
The algorithm proceeds first by computing the system $\widetilde{DS^{2}}$\
from $\widetilde{DS^{0}}$\ and $\widetilde{D_{c}^{0}},$ then it determines
the system $\widetilde{D_{c}^{2}}$.

{\scriptsize {$\widetilde{E_{c}^{2}}$\textit{=}$\left( 
\begin{tabular}{lll}
\multicolumn{3}{l}{%
\begin{tabular}{ll}
$M^{2}:p_{2},p_{3},p_{3},p_{7}\rightarrow 1$ & $Ni^{2}$: $%
\{t_{1},t_{2},t_{7}\}\rightarrow -1;t_{3}\rightarrow 2.$ \\ 
$Ne^{2}$: $\{t_{1},t_{3}\}\rightarrow 0;\{t_{2},t_{7}\}\rightarrow 2.$ & $%
Na^{2}$: $\{t_{7},t_{2}\}\rightarrow 2;t_{3}\rightarrow 0;t_{1}\rightarrow
0. $%
\end{tabular}%
} \\ 
\begin{tabular}{|c|c|c|c|c|}
\hline
$\widetilde{D_{c}^{2}}$ & $t_{1}$ & $t_{2}$ & $t_{3}$ & $t_{7}$ \\ \hline
$t_{1}$ & 0 & 4 & 1 & 9 \\ \hline
$t_{2}$ & 1 & 0 & 2 & 8 \\ \hline
$t_{3}$ & 1 & 5 & 0 & 10 \\ \hline
$t_{7}$ & -7 & -5 & -6 & 0 \\ \hline
\end{tabular}
& 
\begin{tabular}{l}
\begin{tabular}{|c|c|c|c|c|}
\hline
$\widetilde{DS^{2}}[i,t]$ & $t_{1}$ & $t_{2}$ & $t_{3}$ & $t_{7}$ \\ \hline
{\tiny 0} & {\tiny 3} & {\tiny 7} & {\tiny 4} & {\tiny 12} \\ \hline
n={\tiny 2} & {\tiny 3} & {\tiny 5} & {\tiny 4} & {\tiny 10} \\ \hline
\end{tabular}%
\medskip \\ 
\begin{tabular}{|c|c|c|c|c|}
\hline
$\widetilde{DS^{2}}[t,i]$ & $t_{1}$ & $t_{2}$ & $t_{3}$ & $t_{7}$ \\ \hline
{\tiny 0} & {\tiny -3} & {\tiny -2} & {\tiny -2} & {\tiny -10} \\ \hline
n={\tiny 2} & {\tiny -1} & {\tiny -2} & {\tiny 0} & {\tiny -10} \\ \hline
\end{tabular}%
\medskip\ 
\end{tabular}%
\medskip & \ 
\begin{tabular}{l}
\begin{tabular}{|c|c|}
\hline
$\widetilde{DS^{2}}[i,n]$ & {\tiny 0} \\ \hline
n=2 & {\tiny 2} \\ \hline
\end{tabular}%
\medskip\  \\ 
\begin{tabular}{|c|c|}
\hline
$\widetilde{DS^{2}}[n,i]$ & {\tiny 0} \\ \hline
n=2 & {\tiny 0} \\ \hline
\end{tabular}%
\medskip\ 
\end{tabular}%
\end{tabular}%
\right. $ } }

Then firing $t_{1}$ from $\widetilde{E_{c}^{2}}$ yields the class $%
\widetilde{E_{c}^{3}}$. At this stage, the resulted \textit{DBM }system%
\textit{\ }$\widetilde{D_{c}^{2}}$ is equal to that obtained in the graph $%
\widetilde{GR}$ after firing the same sequence.

{\scriptsize {$\widetilde{E_{c}^{3}}$\textit{=}$\left( 
\begin{tabular}{lll}
\multicolumn{3}{l}{%
\begin{tabular}{ll}
$M^{3}:p_{2},p_{3},p_{5},p_{7}\rightarrow 1$ & $Ni^{3}$: $%
\{t_{2},t_{5},t_{7}\}\rightarrow -1;t_{3}\rightarrow 2.$ \\ 
$Ne^{3}$: $\{t_{2},t_{7}\}\rightarrow 2;t_{3}\rightarrow 0;t_{5}\rightarrow
3.$ & $Na^{3}$: $\{t_{7},t_{2}\}\rightarrow 2;t_{3}\rightarrow
0;t_{5}\rightarrow 3.$%
\end{tabular}%
} \\ 
\begin{tabular}{|c|c|c|c|c|}
\hline
$\widetilde{D_{c}^{3}}$ & $t_{2}$ & $t_{3}$ & $t_{5}$ & $t_{7}$ \\ \hline
$t_{2}$ & 0 & 4 & 0 & 8 \\ \hline
$t_{3}$ & 4 & 0 & 0 & 9 \\ \hline
$t_{5}$ & 4 & 4 & 0 & 9 \\ \hline
$t_{7}$ & -5 & -3 & -7 & 0 \\ \hline
\end{tabular}%
\medskip & 
\begin{tabular}{l}
\begin{tabular}{|c|c|c|c|c|}
\hline
$\widetilde{DS^{3}}[i,t]$ & $t_{2}$ & $t_{3}$ & $t_{5}$ & $t_{7}$ \\ \hline
{\tiny 0} & {\tiny 7} & {\tiny 7} & {\tiny 3} & {\tiny 12} \\ \hline
{\tiny 2} & {\tiny 5} & {\tiny 7} & {\tiny 3} & {\tiny 10} \\ \hline
n={\tiny 3} & {\tiny 4} & {\tiny 4} & {\tiny 0} & {\tiny 9} \\ \hline
\end{tabular}%
\medskip \\ 
\begin{tabular}{|c|c|c|c|c|}
\hline
$\widetilde{DS^{3}}[t,i]$ & $t_{2}$ & $t_{3}$ & $t_{5}$ & $t_{7}$ \\ \hline
{\tiny 0} & {\tiny -3} & {\tiny -3} & {\tiny -3} & {\tiny -10} \\ \hline
{\tiny 2} & {\tiny -2} & {\tiny -1} & {\tiny -1} & {\tiny -10} \\ \hline
n={\tiny 3} & {\tiny 0} & {\tiny 0} & {\tiny 0} & {\tiny -7} \\ \hline
\end{tabular}%
\medskip%
\end{tabular}%
\  & \ 
\begin{tabular}{l}
\begin{tabular}{|c|c|c|}
\hline
$\widetilde{DS^{3}}[i,n]$ & {\tiny 0} & {\tiny 2} \\ \hline
n=3 & {\tiny 3} & {\tiny 3} \\ \hline
\end{tabular}%
\medskip\  \\ 
\begin{tabular}{|c|c|c|}
\hline
$\widetilde{DS^{3}}[n,i]$ & {\tiny 0} & {\tiny 2} \\ \hline
n=3 & {\tiny -3} & {\tiny -1} \\ \hline
\end{tabular}%
\medskip\ 
\end{tabular}%
\end{tabular}%
\right. $ }}

Firing the transition $t_{2}$ from the previous class leads to $\widetilde{%
E_{c}^{5}}$. Here, we notice that the minimal residual time of the
persistent inhibited transition $t_{3}$ relatively to the point $(0)$ has
increased to 4, as we have $\widetilde{DS^{5}}[t_{3},0]=-4.$ {\scriptsize {$%
\widetilde{\mathit{E}_{\mathit{c}}^{5}}$\textit{=}$\left( 
\begin{tabular}{lll}
\multicolumn{3}{l}{%
\begin{tabular}{ll}
$M^{5}:p_{3},p_{5},p_{7}\rightarrow 1$ & $Ni^{5}$: $\{t_{5},t_{7}\}%
\rightarrow -1;t_{3}\rightarrow 2.$ \\ 
$Ne^{5}$: $t_{7}\rightarrow 2;t_{3}\rightarrow 0;t_{5}\rightarrow 3.$ & $%
Na^{5}$: $t_{7}\rightarrow 2;t_{3}\rightarrow 0;t_{5}\rightarrow 3.$%
\end{tabular}%
} \\ 
\begin{tabular}{|c|c|c|c|}
\hline
$\widetilde{D_{c}^{5}}$ & $t_{3}$ & $t_{5}$ & $t_{7}$ \\ \hline
$t_{3}$ & 0 & -1 & 7 \\ \hline
$t_{5}$ & 4 & 0 & 8 \\ \hline
$t_{7}$ & -3 & -7 & 0 \\ \hline
\end{tabular}%
\medskip & \medskip 
\begin{tabular}{l}
\begin{tabular}{|c|c|c|c|}
\hline
$\widetilde{DS^{5}}[i,t]$ & $t_{3}$ & $t_{5}$ & $t_{7}$ \\ \hline
{\tiny 0} & {\tiny 7} & {\tiny 3} & {\tiny 11} \\ \hline
{\tiny 2} & {\tiny 7} & {\tiny 3} & {\tiny 10} \\ \hline
3 & 4 & 0 & 8 \\ \hline
n={\tiny 5} & {\tiny 4} & {\tiny 0} & {\tiny 8} \\ \hline
\end{tabular}%
\smallskip \\ 
\begin{tabular}{|c|c|c|c|}
\hline
$\widetilde{DS^{5}}[t,i]$ & $t_{3}$ & $t_{5}$ & $t_{7}$ \\ \hline
{\tiny 0} & {\tiny -4} & {\tiny -3} & {\tiny -10} \\ \hline
{\tiny 2} & {\tiny -2} & {\tiny -2} & {\tiny -10} \\ \hline
3 & 0 & 0 & -7 \\ \hline
n={\tiny 5} & {\tiny -1} & {\tiny 0} & {\tiny -7} \\ \hline
\end{tabular}%
\medskip\ 
\end{tabular}
& \ 
\begin{tabular}{l}
\begin{tabular}{|c|c|c|c|}
\hline
$\widetilde{DS^{5}}[i,n]$ & {\tiny 0} & {\tiny 2} & 3 \\ \hline
n=5 & {\tiny 3} & {\tiny 3} & 0 \\ \hline
\end{tabular}%
\medskip\  \\ 
\begin{tabular}{|c|c|c|c|}
\hline
$\widetilde{DS^{5}}[n,i]$ & {\tiny 0} & {\tiny 2} & 3 \\ \hline
n=5 & {\tiny -3} & {\tiny -2} & 0 \\ \hline
\end{tabular}%
\medskip\ 
\end{tabular}%
\end{tabular}%
\right. $}}

The formula given in Proposition 1 suggests to compute this distance from
the system $\widetilde{DS^{2}}$, since $2$ is the point that inhibited $%
t_{3} $ for the last time. As $t_{3}$ still remains inhibited in the
sequence we have according to Proposition 1, $\widetilde{DS^{5}}%
[t_{3},0]=MIN(\widetilde{DS^{2}}[t_{3},0]+\widetilde{DS^{5}}[2,5],$ $\ $ \ $%
\widetilde{DS^{3}}[t_{3},0]+\widetilde{DS^{3}}[t_{2},3])$; we obtain $%
\widetilde{DS^{5}}[t_{3},0]=MIN(-2-2,$ \ $-3+0)=-4$. Hence we compute the
minimal residual time of $t_{3}$ relatively to point n=5 and we obtain: $%
\widetilde{DS^{5}}[t_{3},5]=${\scriptsize {{\small $-1.$} } }

Comparatively to the construction of the graph $\widetilde{GR}$, this class
is better approximated in $\widetilde{GRc}.$ This prevents the appearance of
false behaviors as it is the case in $\widetilde{GR}.$ To highlight this
fact, let us consider the firing of the transition $t_{5}$ from $\widetilde{%
\mathit{E}_{\mathit{c}}^{5}}$ which produces the class $\widetilde{\mathit{E}%
_{\mathit{c}}^{6}}.${\small \ }

{\scriptsize {$\widetilde{\mathit{E}_{\mathit{c}}^{6}}$\textit{=}$\left( 
\begin{tabular}{lll}
\multicolumn{3}{l}{%
\begin{tabular}{ll}
$M^{6}:p_{3},p_{6}\rightarrow 1$ & $Ni^{6}$: $t_{6}\rightarrow
-1;t_{3}\rightarrow 2.$ \\ 
$Ne^{6}$: $t_{3}\rightarrow 0;t_{6}\rightarrow 6.$ & $Na^{6}$:$%
\{t_{6},t_{3}\}\rightarrow 6.$%
\end{tabular}%
} \\ 
\begin{tabular}{|c|c|c|}
\hline
$\widetilde{D_{c}^{6}}$ & $t_{3}$ & $t_{6}$ \\ \hline
$t_{3}$ & 0 & -1 \\ \hline
$t_{6}$ & 4 & 0 \\ \hline
\end{tabular}%
\medskip & \medskip 
\begin{tabular}{ll}
\begin{tabular}{|c|c|c|}
\hline
$\widetilde{DS^{6}}[i,t]$ & $t_{3}$ & $t_{6}$ \\ \hline
{\tiny 0} & {\tiny 7} & {\tiny 3} \\ \hline
2 & 7 & 3 \\ \hline
n={\tiny 6} & {\tiny 4} & {\tiny 0} \\ \hline
\end{tabular}%
\smallskip & 
\begin{tabular}{|c|c|c|}
\hline
$\widetilde{DS^{6}}[t,i]$ & $t_{3}$ & $t_{6}$ \\ \hline
{\tiny 0} & {\tiny -4} & {\tiny -3} \\ \hline
2 & -2 & -2 \\ \hline
n={\tiny 6} & {\tiny -1} & {\tiny 0} \\ \hline
\end{tabular}%
\medskip%
\end{tabular}
& \ 
\begin{tabular}{l}
\begin{tabular}{|c|c|c|}
\hline
$\widetilde{DS^{6}}[i,n]$ & {\tiny 0} & 2 \\ \hline
n=6 & {\tiny 3} & 3 \\ \hline
\end{tabular}%
\medskip\  \\ 
\begin{tabular}{|c|c|c|}
\hline
$\widetilde{DS^{6}}[n,i]$ & {\tiny 0} & 2 \\ \hline
n=6 & {\tiny -3} & -2 \\ \hline
\end{tabular}%
\medskip\ 
\end{tabular}%
\end{tabular}%
\right. $} }

As we notice, $\widetilde{\mathit{E}_{\mathit{c}}^{6}}$ is exactly
approximated relatively to the exact class obtained after firing the same
sequence in the graph $GR$. Only the transition $t_{6}$ is firable from $%
\widetilde{\mathit{E}_{\mathit{c}}^{6}}$ whereas both $t_{6}$ and $t_{3}$
are firable from the corresponding class in the graph $\widetilde{GR}$. The
same observation is made when considering the alternative firing sequence $%
\widetilde{E_{c}^{0}}\overset{t_{4}}{\leadsto }\widetilde{E_{c}^{2}}\overset{%
t_{1}}{\leadsto }$ $\widetilde{E_{c}^{3}}\overset{t_{5}}{\leadsto }$ $%
\widetilde{E_{c}^{9}}\overset{t_{2}}{\leadsto }\widetilde{E_{c}^{6}},$ . In
this sequence, the transition $t_{3}$ is no longer inhibited when reaching
the class $\widetilde{E_{c}^{9}\text{ }}$ and we have:

{\scriptsize {$\widetilde{\mathit{E}_{\mathit{c}}^{9}}$\textit{=}$\left( 
\begin{tabular}{lll}
\multicolumn{3}{l}{%
\begin{tabular}{ll}
$M^{9}:p_{3},p_{2},p_{6}\rightarrow 1$ & $Ni^{9}$: $t_{6},t_{2}\rightarrow
-1;t_{3}\rightarrow 2.$ \\ 
$Ne^{9}$: $t_{3}\rightarrow 0;t_{2}\rightarrow 2;t_{6}\rightarrow 6.$ & $%
Na^{9}$:$\{t_{6},t_{3}\}\rightarrow 9;t_{2}\rightarrow 2.$%
\end{tabular}%
} \\ 
\begin{tabular}{|c|c|c|c|}
\hline
$\widetilde{D_{c}^{9}}$ & $t_{2}$ & $t_{3}$ & $t_{6}$ \\ \hline
$t_{2}$ & {\tiny 0} & {\tiny 4} & {\tiny 0} \\ \hline
$t_{3}$ & {\tiny 4} & {\tiny 0} & {\tiny 0} \\ \hline
$t_{6}$ & {\tiny 4} & {\tiny 4} & {\tiny 0} \\ \hline
\end{tabular}%
\medskip & \medskip 
\begin{tabular}{ll}
\begin{tabular}{|c|c|c|c|}
\hline
$\widetilde{DS^{9}}[i,t]$ & $t_{2}$ & $t_{3}$ & $t_{6}$ \\ \hline
{\tiny 0} & {\tiny 7} & {\tiny 7} & {\tiny 3} \\ \hline
{\tiny 2} & {\tiny 5} & {\tiny 7} & {\tiny 3} \\ \hline
n={\tiny 9} & {\tiny 4} & {\tiny 4} & {\tiny 0} \\ \hline
\end{tabular}%
\smallskip & 
\begin{tabular}{|c|c|c|c|}
\hline
$\widetilde{DS^{9}}[t,i]$ & $t_{2}$ & $t_{3}$ & $t_{6}$ \\ \hline
{\tiny 0} & {\tiny -3} & {\tiny -3} & {\tiny -3} \\ \hline
2 & {\tiny -2} & {\tiny -1} & {\tiny -1} \\ \hline
n={\tiny 6} & {\tiny 0} & {\tiny 0} & {\tiny 0} \\ \hline
\end{tabular}%
\medskip%
\end{tabular}
& 
\begin{tabular}{l}
\begin{tabular}{|c|c|c|}
\hline
$\widetilde{DS^{9}}[i,n]$ & {\tiny 0} & {\tiny 2} \\ \hline
n=9 & {\tiny 3} & {\tiny 3} \\ \hline
\end{tabular}%
\medskip\  \\ 
\begin{tabular}{|c|c|c|}
\hline
$\widetilde{DS^{9}}[n,i]$ & {\tiny 0} & {\tiny 2} \\ \hline
n=9 & {\tiny -3} & {\tiny --1} \\ \hline
\end{tabular}%
\medskip\ 
\end{tabular}%
\end{tabular}%
\right. $} }

At this stage, the minimal residual time of the activated transition $t_{3}$
is equal to 0, but it increases to 1 after firing $t_{2}$ to reach the class 
$\widetilde{E_{c}^{6}}.$ Indeed, the firing of $t_{2}$ restricts the space
to the states that have fired initially $t_{4}$ between [0,1]. The formula
provided in Proposition 1 allows to restore appropriate time information to
exactly approximate the reachable class. Therefore, as the inhibition of $%
t_{3}$ occurs and stops earlier in the sequence, the formula suggests to
recompute the inhibition time of $t_{3}$ to better approximate the
calculation of its residual times. Hence, we find that the distance $%
\widetilde{DS^{6}}[t_{3},0]$ has decreased from -3 to -4, and therefore we
obtain $\widetilde{DS^{6}}[t_{3},6]=-1.$

If we consider now the sequence $\widetilde{E_{c}^{0}}\overset{t_{4}}{%
\leadsto }\widetilde{E_{c}^{2}}\overset{t_{1}}{\leadsto }$ $\widetilde{%
E_{c}^{3}}\overset{t_{5}}{\leadsto }$ $\widetilde{E_{c}^{9}}\overset{t_{3}}{%
\leadsto }\widetilde{E_{c}^{10}}$, we notice that $t_{2}$ is fired from $%
\widetilde{E_{c}^{10}}$ whereas it is not in the exact graph $GR$ from the
class $E^{4}$. In actual fact, as $t_{3}$ is fired before $t_{2}$, some of
the points connected to $t_{3}$ are no longer stored in the class $%
\widetilde{E_{c}^{10}}.$ Therefore, the time information computed in the
class $\widetilde{E_{c}^{9}}$ that could better approximate the class $%
\widetilde{E_{c}^{10}}$ is removed after firing $t_{3}$. As a result, the
sequence $\widetilde{E_{c}^{10}}\overset{t_{2}}{\leadsto }$ $\widetilde{%
E_{c}^{14}}\overset{t_{6}}{\leadsto }\widetilde{E_{c}^{8}\text{ }}$is added
mistakenly in both graphs $\widetilde{GRc}$ and $\widetilde{GR}.$

In other respects, we notice that the amount of data needed to represent
each class of the graph $\widetilde{GRc}$ is two or three times higher than
in the graph $\widetilde{GR}$. However, this additional data may be very
useful for the time analysis of the model as it makes it possible, for
instance, to determine efficiently the quantitative properties of the model.
Therefore, no further greedy computation is needed for such a process,
unlike other graph constructions which require to perform further
calculations to determine such properties. For example, in \cite{Bert-grid}%
\cite{IHTPN} the authors have proposed to extend the original model with an
observer containing additional places and transitions modeling the
quantitative property to determine. This method is quite costly as it
requires to compute the reachability graph of the extended model for each
value of the quantitative property to check. Therefore, many graphs
generations may be necessary to determine the exact value of the
quantitative property. In \cite{Bucci}, the authors proposed an interesting
method for quantitative timed analysis. They compute first the $DBM$
overapproximation of the graph. Then, given an untimed transition sequence
from the resulted graph, they can obtain the feasible timings of the
sequence as the solution of a linear programming problem. In particular, if
there is no solution, the sequence has been introduced by the
overapproximation and can be cleaned up, otherwise the solution set allows
to determine the quantitative properties of the considered sequence.
However, this method consumes for each sequence to handle an exponential
complexity time, as a result of solving a general linear programming problem.

As regards our graph construction, the quantitative properties can be
extracted from the graph $\widetilde{GRc}$ in almost all cases without
further computations \cite{Abdelli un}. In the other cases, we need to
perform small calculations. So, let us consider a firing sequence $%
S=(t^{i+1},..,t^{n})$; $S$ describes a path in the graph going from the node
representing the class $\widetilde{E^{i}}$ to the node which represents the
class $\widetilde{E^{n}}$. The system $\widetilde{DS^{n}}$ provides the
minimal and the maximal time distances from transition's enabling,
activating and inhibiting points to point $(n)$ . Hence, to measure the
minimal or the maximal times of the sequence $(t_{f}^{i+1},..,t_{f}^{n})$,
we need to check whether the elements $\widetilde{DS^{n}}[i,n]$ and $%
\widetilde{DS^{n}}[n,i]$ are already computed in the system $\widetilde{%
DS^{n}}$ namely that $i\in Point^{n}$. Otherwise, if $(i)$ does not belong
to the latter, then we need to perform further computations on the final
graph by using $Proposition.1.$ In concrete terms, the idea is to extend the
set of points with the missing point $(i),$ and this for every node $(j)$ of
the path going from the node $(i+1)$ to the node $(n-1)$. Then, we compute
in the systems $\widetilde{DS^{j}}$ only the time distances involving the
missing point $(i)$ since the other distances are already computed when
generating the graph. The process carries on until reaching the node $(n);$
there we should have achieved the computation of $DS^{n}[n,i]$ and $%
DS^{n}[i,n]$.

To determine the $BCRT$ and the $WCRT$\ (best and worst cases response
times), of a task, we should repeat this process for all the related
sequences in the graph. For this effect, the graph is first computed to
browse for the sequences to handle. Then each sequence is analyzed to
extract its quantitative properties.

In order to advocate the efficiency of our graph construction, we give in
the next section some experimental results that compare the performances of
our algorithms with those of other approaches.

\subsection{Experimental results}

We have implemented ours algorithms using $C++$ builder language on a
Windows XP workstation. The experiments have been performed on a Pentium $V$
with a processor speed of $2,27$ $GH$ and $2$ $GB$ of RAM capacity. The
different tests have been carried out while using four tools: $ORIS$ tool 
\cite{ORIS TOOL}, $\mathit{TINA}$\textit{\ }tool \cite{tina}, $\mathit{ROMEO}
$ tool \cite{Romeo}, and our tool named $ITPN$\textit{\ Analyzer \cite{ITPN}}%
. Then, we compared the obtained graphs while considering three parameters,
the number of classes, the number of edges and the computation times.
Thereafter, we denote by $\mathit{NF}$ (\textit{Not Finished}) the tests
that have spent more than 5 minutes of time computation or that have led to
memory overflows. Moreover, we denote by $\mathit{NA}$ (\textit{Not Available%
}) when no parameter measurement is provided by the tool.

\begin{figure}[tbph]
\centering\includegraphics[width=8 cm]{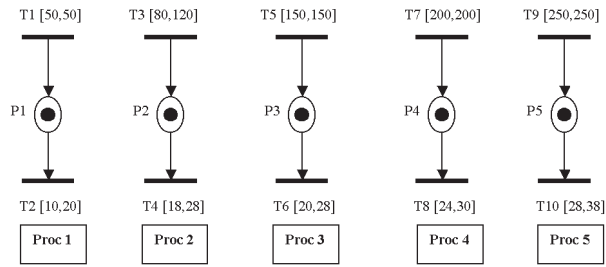} \centering
\caption{$TPN$ used in the experiments.}
\end{figure}

\begin{table}[th]
\caption{Results of experiments performed on $TPN$\textit{.}}\centering%
{\tiny {\ {\scriptsize {\ 
\begin{tabular}{|c|c|c|c|c|c|c|}
\hline
Examples & Tools & TINA & ROMEO & ORIS & \multicolumn{2}{|c|}{ITPNAnalyzer}
\\ \hline
&  & DBM & DBM & DBM & DBM & Tdis \\ \hline
& Classes & 2 & 2 & 2 & 2 & 2 \\ \cline{2-7}\cline{5-7}
Proc 1 & Edges & 2 & 2 & $NA$ & 2 & 2 \\ \cline{2-7}\cline{5-7}
& Times (ms) & 0 & $NA$ & 0 & 0 & 0 \\ \hline
& Classes & 186 & 186 & 188 & 186 & 186 \\ \cline{2-7}\cline{5-7}
Proc 1 2 & Edges & 262 & 262 & $NA$ & 262 & 262 \\ \cline{2-7}\cline{5-7}
& Times (ms) & 0 & $NA$ & 16 & 0 & 0 \\ \hline
& Classes & 958 & 958 & 1.038 & 958 & 958 \\ \cline{2-7}\cline{5-7}
Proc 1 2 3 & Edges & 1506 & 1506 & $NA$ & 1506 & 1506 \\ 
\cline{2-7}\cline{5-7}
& Times (ms) & 16 & $NA$ & 391 & 5 & 30 \\ \hline
& Classes & 5.219 & 5.219 & 6.029 & 5.219 & 5.219 \\ \cline{2-7}\cline{5-7}
Proc 1 2 3 4 & Edges & 8.580 & 8.580 & $NA$ & 8.580 & 8.580 \\ 
\cline{2-7}\cline{5-7}
& Times (ms) & 140 & $NA$ & 2.719 & 38 & 80 \\ \hline
& Classes & 42.909 & 42.909 & 52.452 & 42.909 & 42.909 \\ 
\cline{2-7}\cline{5-7}
Proc 1 2 3 4 5 & Edges & 73.842 & 73.842 & $NA$ & 73.842 & 73.842 \\ 
\cline{2-7}\cline{5-7}
& Times (ms) & 6.739 & $NA$ & 30.000 & 670 & 1.310 \\ \hline
\end{tabular}%
} }}}
\end{table}

The first tests that have been carried out intended to verify whether our $%
\mathit{TPN}$\ graph construction produce the same graphs as when using
other tools. For this effect, we have considered the combination of the $%
\mathit{TPNs}$\ shown in $\mathit{Fig.6}$. First, we started by testing the
net \textit{Proc1,} then we composed \textit{Proc1} with \textit{Proc2}, \
and so on. The results of these experiments are reported in $\mathit{Table}$ 
$\mathit{3}$\textit{.} The latter shows that all graphs are identical except
for ORIS which extends the expression of a class to the parameter $NEW$%
\footnote{$NEW(t)$ is a boolean that denotes whether the transition $t$ is
newly enabled or not. Therefore, although they are bisimilar, two classes
that have the same marking $M$ and the same firing space $D$ are considered
as non equivalent if the parameter $New$ is not identical in both classes.}.
In other respects, as it was expected, the times needed to compute the 
\textit{DBM} state class graphs are faster than when computing the time
distance based graphs.

The second series of tests that have been performed aimed at comparing the
construction of the graph $\widetilde{GRc}$ with other graph construction
approaches. For this effect, we considered the $ITPN$ model shown in \textit{%
Fig.7} while varying the intervals of transitions \textit{t}$_{2},$ and 
\textit{t}$_{3}$. The results of these tests are reported in $\mathit{Table}$
$\mathit{4}$. This $ITPN,$\ presented previously in \cite{Bucci}, describes
three independent tasks that are in conflict for a common resource (e.g 
\textit{a processor}), and given respectively by the following pairs of
transitions: Task$_{1}$=$(t_{1},t_{4})$, Task$_{2}$=$(t_{2},t_{5})$ and Task$%
_{3}$=$(t_{3},t_{6})$. Task 1 has a higher priority than the two other
tasks, and Task 2 has priority on the Task 3. The priorities are
characterized by using inhibitors arcs that connect the place $p_{1}$ to the
transitions $t_{5}$ and $t_{6};$ and\ the place $p_{2}$ to the transition $%
t_{6}.$

\begin{figure}[h]
\centering\includegraphics[width=6 cm]{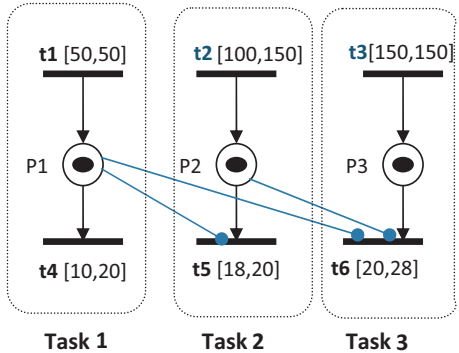}
\caption{An \textit{ITPN}\ example modeling three conflicting tasks.}
\end{figure}

For this purpose, different approaches have been tested: The exact graph
construction defined in \cite{LIme2003} and its \textit{DBM}
overapproximation defined in \cite{IHTPN} which are both implemented in $%
\mathit{ROMEO}$; the \textit{DBM} overapproximation defined in \cite{abdelli}
and the time distance based approximation defined in this paper which are
both implemented in $\mathit{ITPN\ Analyzer}$; the \textit{DBM}
overapproximation defined in \cite{Bucci} and implemented in $\mathit{ORIS}$%
; and finally, the K-grid based approximation defined in \cite{Bert-grid}
and implemented in $TINA$. Notice that for the latter construction, we have
considered the highest grid to approximate the polyhedra. In this case, this
approach succeeds to compute the exact graph in almost all cases, but
nevertheless with the highest cost.

\begin{table}[tbph]
\caption{Results of experiments performed on $\mathit{ITPN}$\textit{.}}%
\centering{\tiny {\ 
\begin{tabular}{|c|c|c|c|c|c|c|c|}
\hline
& TOOLS & TINA & \multicolumn{2}{|c|}{ROMEO} & \multicolumn{2}{|c|}{ITPN
Analyser} & ORIS \\ \cline{2-8}\cline{3-8}
Examples & Methods & K-grid & Exact & DBM & $\widetilde{GR}(DBM)$ & $%
\widetilde{GRc}(TDis)$ & DBM \\ \hline\hline
t$_{2}$\ [100,150] & Classes & 4.489 & 4.489 & 5.431 & 5.378 & 4.483 & 5.538
\\ \cline{2-8}\cline{6-8}
t$_{3}$\ [160,160] & Edges & 6.360 & 6.360 & 7.608 & 7.530 & 6.345 & NA \\ 
\cline{2-8}\cline{6-8}
& Times(ms) & 1632 & NA & NA & 60 & 80 & 1578 \\ \hline\hline
t$_{2}$\ [80,120] & Classes & 27.901 & NF & \ 47.777 & 39.648 & 27.889 & 
40.414 \\ \cline{2-8}\cline{2-8}
t$_{3}$\ [145,145] & Edges & 40.073 & NF & 67.754 & 56.238 & 40.163 & NA \\ 
\cline{2-8}\cline{2-8}
& Times(ms) & 5.086 & NA & NA & 300 & 530 & 5.360 \\ \hline\hline
t$_{2}$\ [80,120] & Classes & 29.976 & NF & \ 47.888 & 42.247 & 29.964 & 
42.733 \\ \cline{2-8}
t$_{3}$\ [165,165] & Edges & 42.844 & NF & 67.546, & 59.635 & 42.832 & NA \\ 
\cline{2-8}
& Times(ms) & 5.522 & NA & NA & 220 & 580 & 5.188 \\ \hline\hline
t$_{2}$\ [100,150] & Classes & 16.913 & 16.913 & 21.033 & 20.802 & 16.901 & 
21.116 \\ \cline{2-8}
t$_{3}$\ [145,145] & Edges & 23.583 & 23.583 & 28.989 & 28.635 & 23.571 & NA
\\ \cline{2-8}
& Times(ms) & 2.870 & NA & NA & 170 & 230 & 5157 \\ \hline\hline
t$_{2}$\ [100,150] & Classes & 320 & 320 & 403 & 394 & 319 & 429 \\ 
\cline{2-8}\cline{6-8}
t$_{3}$\ [150,150] & Edges & 460 & 460 & 575 & 562 & 459 & NA \\ 
\cline{2-8}\cline{6-8}
& Times(ms) & 110 & NA & NA & 4 & 10 & 156 \\ \hline\hline
t$_{2}$\ [100,150] & Classes & 4.142 & 4.142 & 5.034 & 4.982 & 4.136 & 5140
\\ \cline{2-8}\cline{6-8}
t$_{3}$\ [140,140] & Edges & 5.889 & 5.889 & 7.095 & 7.014 & 5.883 & NA \\ 
\cline{2-8}\cline{6-8}
& Times(ms) & 1502 & NA & NA & 20 & 80 & 1.765 \\ \hline\hline
t$_{3}$\ [80,120] & Classes & 28 392 & NF & 47.622 & 40.842 & 28.920 & 41.368
\\ \cline{2-8}\cline{6-8}
t$_{5}$\ [155,155] & edges & 41.452 & NF & 67.309 & 57.766 & 41.440 & NA \\ 
\cline{2-8}\cline{6-8}
& Times (ms) & 15703 & NA & NA & 280 & 640 & 9.062 \\ \hline\hline
t$_{2}$\ [80,120] & Classes & 7.018 & 7.018 & 12.379 & 10.004 & 7.012 & 
10.400 \\ \cline{2-8}
t$_{3}$\ [140,140] & Edges & 10.242 & 10.242 & 17.829 & 14.406 & 10.236 & NA
\\ \cline{2-8}
& Times(ms) & 2834 & NA & NA & 50 & 140 & 3.516 \\ \hline\hline
t$_{2}$\ [100,150] & Classes & 11.351 & 11.351 & 16.354 & 15.178 & 11.339 & 
15.318 \\ \cline{2-8}
t$_{3}$\ [135,135] & Edges & 15.649 & 15.649 & 22.230 & 20.486 & 15.639 & NA
\\ \cline{2-8}
& Times(ms) & 4907 & NA & NA & 90 & 230 & 4765 \\ \hline\hline
t$_{2}$\ [100,150] & Classes & 17.612 & 17.612 & 21.857 & 21.626 & 17.600 & 
21.942 \\ \cline{2-8}
t$_{3}$\ [155,155] & Edges & 24.522 & 24.522 & 30.065 & 29.711 & 24.520 & NA
\\ \cline{2-8}
& Times(ms) & 7951 & NA & NA & 130 & 230 & 5594 \\ \hline
\end{tabular}%
} }
\end{table}

As we notice, the graphs computed by the considered \textit{DBM}
overapproximations are not identical. As concerns $ORIS,$ the reason is
given above.However for \textit{ROMEO}, we have shown in \cite{abdelli} that
the $DBM$ approximation defined in \cite{IHTPN} is not truly implemented. In
actual fact, in \textit{ROMEO} the normalization of the \textit{DBM} system
is performed after removing the polyhedral inequalities, whereas it must be
done before, thus yielding a loss of precision in the resulted graphs. It is
noteworthy that among these \textit{DBM} overapproximations, the approach
defined in \cite{abdelli} and implemented in \cite{ITPN} is the one that
computes the tightest graphs with the fastest times. Concerning the time
distance based approximation defined in this paper, the results show that
the obtained graphs are of the sime size relatively to the exact ones, even
smaller. Moreover, the times needed for their computation are 10 even 20
times faster than those of \textit{TINA}, and slightly more comparing to 
\textit{ROMEO}.

It should be noticed that although the $\widetilde{GRc}$ is almost equal to $%
GR$, it still remains an overapproximation of it. In actual fact, many
classes that stand unequal in $GR$ despite they are bisimilar, become
equivalent when they are approximated\footnote{%
The polyhedral inequalities that prevent class' equality in $GR$ and hence
their equivalence are removed in $\widetilde{GRc}$.} in $\widetilde{GRc},$
thus compacting its size comparatively to the graph $GR$.

\begin{table}[tbph]
\caption{WCRT and BCRT estimation of Task 3.}\centering\centering{\tiny {\ 
\begin{tabular}{|c||c|c|c|c|}
\hline
{\scriptsize Examples} & Methods & $\widetilde{GR}(DBM)$ & $\widetilde{GRc}%
(Tdis)$ & {\scriptsize Exact} \\ \hline\hline
{\scriptsize t}$_{2}${\scriptsize \ [100,150]} & {\scriptsize WCRT} & 
{\scriptsize 126} & {\scriptsize 88} & {\scriptsize 88} \\ 
\cline{2-5}\cline{2-5}
{\scriptsize t}$_{3}${\scriptsize \ [160,160]} & {\scriptsize BCRT} & 
{\scriptsize 20} & {\scriptsize 20} & {\scriptsize 20} \\ \hline\hline
{\scriptsize t}$_{2}${\scriptsize \ [100,150]} & {\scriptsize WCRT} & 
{\scriptsize 126} & {\scriptsize 88} & {\scriptsize 88} \\ 
\cline{2-5}\cline{2-5}
{\scriptsize t}$_{3}${\scriptsize \ [150,150]} & {\scriptsize BCRT} & 
{\scriptsize 30} & {\scriptsize 30} & {\scriptsize 30} \\ \hline\hline
{\scriptsize t}$_{2}${\scriptsize \ [80,120]} & {\scriptsize WCRT} & 
{\scriptsize 198} & {\scriptsize 128} & {\scriptsize 128} \\ \cline{2-5}
{\scriptsize t}$_{3}${\scriptsize \ [130,130]} & {\scriptsize BCRT} & 
{\scriptsize 20} & {\scriptsize 20} & {\scriptsize 20} \\ \hline\hline
{\scriptsize t}$_{2}${\scriptsize \ [100,150]} & {\scriptsize WCRT} & 
{\scriptsize 126} & {\scriptsize 88} & {\scriptsize 88} \\ \cline{2-5}
{\scriptsize t}$_{3}${\scriptsize \ [135,135]} & {\scriptsize BCRT} & 
{\scriptsize 20} & {\scriptsize 20} & {\scriptsize 20} \\ \hline\hline
{\scriptsize t}$_{2}${\scriptsize \ [100,150]} & {\scriptsize WCRT} & 
{\scriptsize 126} & {\scriptsize 88} & {\scriptsize 88} \\ \cline{2-5}
{\scriptsize t}$_{3}${\scriptsize \ [155,155]} & {\scriptsize BCRT} & 
{\scriptsize 20} & {\scriptsize 20} & {\scriptsize 20} \\ \hline\hline
{\scriptsize t}$_{2}${\scriptsize \ [80,120]} & {\scriptsize WCRT} & 
{\scriptsize 208} & {\scriptsize 128} & {\scriptsize 128} \\ \cline{2-5}
{\scriptsize t}$_{3}${\scriptsize \ [140,140]} & {\scriptsize BCRT} & 
{\scriptsize 20} & {\scriptsize 20} & {\scriptsize 20} \\ \hline
\end{tabular}
}}
\end{table}

The final tests, results of which are given in $\mathit{Tab}$ $\mathit{5,}$
report the obtained \textit{BCRT} and \textit{WCRT} of Task 3, while
assuming different graph constructions. As we can see, the computed values
of the \textit{BCRT} are the same for all the nets whatever the graph
construction we consider. On the other hand, the \textit{WCRT} is
differently estimated following the approach we use. When considering the
graph $\widetilde{GR}$, the approximated values are too coarse as it is the
case in the tests 3 and 6. However, the $\widetilde{GRc}$ preserves the
exact value of the \textit{WCRT}\ for all the tested nets. These results
show how tight this approximation is, because the additional sequences that
are distorting the estimation of the \textit{WCRT} in the graph $\widetilde{%
GR}$ are completely removed in the $\widetilde{GRc}$.

\section{Conclusion}

We have proposed in this paper a novel approach to compute an
overapproximation of the state space of real time preemptive systems modeled
using the $ITPN$ model. For this effect, we have defined the time distance
system that encodes the quantitative properties of each class of states
reachable in the exact graph. Then we have provided efficient algorithms to
overapproximate its coefficients and to compute the \textit{DBM}
overapproximation of a class. We proved that this construction is more
precise than other classical \textit{DBM} overapproximation defined in the
literature \cite{abdelli} \cite{IHTPN}\cite{Bucci}, and showed how it is
appropriate to restore the quantitative properties of the model. Simulation
results comparing the performances of our graph construction with other
techniques were reported.

\section{APPENDIX A : Proof of Theorem 1}

{\small {\ We have to determine the following clauses: }}

\begin{enumerate}
\item {\small $\left\rceil D^{0}\right\lceil =\left\rceil \widetilde{D^{0}}%
\right\lceil =\left\rceil \widetilde{D_{c}^{0}}\right\lceil $ and $%
\left\rceil DS^{0}\right\lceil =\left\rceil \widetilde{DS^{0}}\right\lceil $}

\item {\small Let be $S=(t_{f}^{1},..,t_{f}^{n});$ if $(M^{0},D^{0})\overset{%
t_{f}^{1}}{\longmapsto }..\overset{t_{f}^{n-1}}{\longmapsto }%
E^{n-1}=(M^{n-1},D^{n-1})$ , \newline
$(M^{0},Ne^{0},Ni^{0},Na^{0},\widetilde{DS^{0}},\widetilde{D_{c}^{0}})%
\overset{t_{f}^{1}}{\leadsto }..\overset{t_{f}^{n-1}}{\leadsto }\widetilde{%
E_{c}^{n-1}}=(M^{n-1},Ne^{n-1},Ni^{n-1},Na^{n-1},\widetilde{DS^{n-1}},%
\widetilde{D_{c}^{n-1}})$, and $(M^{0},\widetilde{D^{0}})\overset{t_{f}^{1}}{%
\rightsquigarrow }..\overset{t_{f}^{n-1}}{\rightsquigarrow }\widetilde{%
E^{n-1}}=(M^{n-1},\widetilde{D^{n-1}})$ such that $\left\rceil
D^{n-1}\right\lceil \subseteq \left\rceil \widetilde{D_{c}^{n-1}}%
\right\lceil \subseteq \left\rceil \widetilde{D^{n-1}}\right\lceil $ and $%
\left\rceil DS^{n-1}\right\lceil \subseteq \left\rceil \widetilde{DS^{n-1}}%
\right\lceil $ $;$ we have : if $E^{n-1}\overset{t_{f}^{n}}{\longmapsto }%
E^{n}=(M^{n},D^{n}),$ then }

\begin{itemize}
\item {\small $\widetilde{E^{n-1}}\overset{t_{f}^{n}}{\rightsquigarrow }%
\widetilde{E^{n}}=(M^{n},\widetilde{D^{n}}),$ }

\item {\small $\widetilde{E_{c}^{n-1}}$$\overset{t_{f}^{n}}{\leadsto }%
\widetilde{E_{c}^{n}}=(M^{n},Ne^{n},Ni^{n},Na^{n},\widetilde{DS^{n}},%
\widetilde{D_{c}^{n}})$ }

\item {\small and we have $\left\rceil D^{n}\right\lceil \subseteq
\left\rceil \widetilde{D_{c}^{n}}\right\lceil \subseteq \left\rceil 
\widetilde{D^{n}}\right\lceil $ and $\left\rceil DS^{n}\right\lceil
\subseteq \left\rceil \widetilde{DS^{n}}\right\lceil $}
\end{itemize}
\end{enumerate}

{\small The clause $(1)$ holds since the system $D$$^{0}$\ is in $DBM$ ; we
have by definition : \ $D^{0}=\widetilde{D_{c}^{0}}=\widetilde{D^{0}}$ and $%
DS^{0}=\widetilde{DS^{0}}.$\ Let us prove now the clause $(2).$ {Let us
assume $D^{n-1}=\widehat{{D^{n-1}}}\wedge $}$\overrightarrow{{D^{n-1}}}.${\
The system }$\overrightarrow{D^{n-1}}${\ denotes the tightest \textit{DBM}
system extracted from the system }${D}^{{n}}$, and is given by all its
normalized \textit{DBM} inequalities, as follows:}

{\small $\left\{ 
\begin{tabular}{l}
$B_{1}:\forall t_{_{1}}\neq t_{_{2}}\in Te(M^{n-1}),\quad \underline{t_{_{2}}%
}-\underline{t_{_{1}}}\leq \overrightarrow{D^{n-1}}[t_{_{1}},t_{_{2}}]\quad $
\\ 
$B_{2}:\forall t\in Te(M^{n-1}),\quad -\overrightarrow{D^{n-1}}[t,\bullet
]\leq \underline{t}\leq \overrightarrow{D^{n-1}}[\bullet ,t]\quad $%
\end{tabular}%
\right. $ }

{\small {According to the hypotheses of the Clause 2, as we have }$%
\left\rceil D^{n-1}\right\lceil \subseteq \left\rceil \overrightarrow{D^{n-1}%
}\right\lceil \subseteq \left\rceil \widetilde{D_{c}^{n-1}}\right\lceil
\subseteq \left\rceil \widetilde{D^{n-1}}\right\lceil $ then the following
properties hold: }

{\small $\left\{ 
\begin{tabular}{l}
$P_{1}:\forall t_{_{1}}\neq t_{_{2}}\in Te(M^{n-1}),\quad \overrightarrow{%
D^{n-1}}[t_{_{1}},t_{_{2}}]\leq \widetilde{D_{c}^{n-1}}[t_{_{1}},t_{_{2}}]%
\leq \widetilde{D^{n-1}}[t_{_{1}},t_{_{2}}].\quad $ \\ 
$P_{2}:\forall t\in Te(M^{n-1}),\quad \overrightarrow{D^{n-1}}[t,\bullet
]\leq \widetilde{D_{c}^{n-1}}[t,\bullet ]\leq \widetilde{D^{n-1}}[t,\bullet
].\quad $ \\ 
$P_{3}:\forall t\in Te(M^{n-1}),\quad \overrightarrow{D^{n-1}}[\bullet
,t]\leq \widetilde{D_{c}^{n-1}}[\bullet ,t]\leq \widetilde{D^{n-1}}[\bullet
,t].\quad $%
\end{tabular}%
\right. $ }

{\small Let us consider now the firing of the transition $t_{f}^{n}$ from $%
E^{n-1}$\ to reach the class $E^{n}$$=(M^{n},D^{n})$$.$ So that $t_{f}^{n}$
can be fired from $E^{n-1}$\ we must have: $t_{f}^{n}\in Ta(M^{n-1})$ and$%
\quad B_{3}:$ $\forall t\in Ta(M^{n-1})\quad 0\leq $\underline{$t_{f}^{n}$}$%
\leq $\underline{$t$}. \ Therefore, if $t_{f}^{n}$ is firable from $E^{n-1},$
then the system $\left\rceil D^{n-1}\wedge B_{3}\right\lceil \neq \emptyset
; $ hence we have $\forall t\in Ta(M^{n-1}),$ $\overrightarrow{D^{n-1}}%
[t_{f},t]\geq 0$. By using the property $(P_{1})$, we deduce that $\forall
t\in Ta(M^{n-1})$ we have $\widetilde{D^{n-1}}[t_{f}^{n},t]\geq 0$\ and $%
\widetilde{D_{c}^{n-1}}[t_{f}^{n},t]\geq 0$. Hence $\widetilde{\beta ^{n}}%
[t_{f}^{n}]\ \geq 0$ and $\widetilde{\beta _{c}^{n}}[t_{f}^{n}]\ \geq 0.$
Consequently, $t_{f}^{n}$ is also firable from the classes }$\widetilde{%
{\small E}_{{\small c}}^{{\small n-1}}}${\small \ and }$\widetilde{{\small E}%
^{{\small n-1}}}.$ {\small It remains to prove that $\left\rceil
DS^{n}\right\lceil \subseteq \left\rceil \widetilde{DS^{n}}\right\lceil $
and }$\left\rceil \overrightarrow{D^{n}}\right\lceil \subseteq \left\rceil 
\widetilde{D_{c}^{n}}\right\lceil \subseteq \left\rceil \widetilde{D^{n}}%
\right\lceil ${\small $.$ \ This requires first to prove that the system $%
\widetilde{DS^{n}}\ $is always an overapproximation of the system }${\small %
DS^{n},}${\small \ namely that each coefficient of $\widetilde{DS^{n}}$ is
equal or greater than its related in }${\small DS^{n-1}}${\small . For this
effect, let us assume the systems }$\widetilde{{\small DS^{n-1}}}${\small \
\ and }${\small DS^{n-1}}${\small \ associated respectively with the class $%
\widetilde{E_{c}^{n-1}}$\ and $E^{n-1}$ given as follows:}

{\small $\left\{ 
\begin{tabular}{l}
$C_{1}:{\Huge \wedge }_{\forall i\in Point^{n-1}}-\widetilde{DS^{n-1}}%
[n-1,i]\leq \underline{t_{f}^{_{i+1}}}+..+\underline{t_{f}^{n-1}}\leq 
\widetilde{DS^{n-1}}[i,n-1]$\newline
\\ 
$C_{2}:${\Huge $\wedge $}$_{\forall t\in Te(M^{n-1})}{\Huge \wedge }%
_{\forall i\in Point^{n-1}\cup \{n-1\}}-\widetilde{DS^{n-1}}[t,i]\leq 
\underline{t_{f}^{_{i+1}}}+..+\underline{t_{f}^{n-1}}+\underline{t}\leq 
\widetilde{DS^{n-1}}[i,t]$%
\end{tabular}%
\newline
\right. $\newline
}

{\small $\left\{ 
\begin{tabular}{l}
$C_{3}:{\Huge \wedge }_{\forall i\in Point^{n-1}}-DS^{n-1}[n-1,i]\leq 
\underline{t_{f}^{_{i+1}}}+..+\underline{t_{f}^{n-1}}\leq DS^{n-1}[i,n-1]$%
\newline
\\ 
$C_{4}:{\Huge \wedge }_{\forall t\in Te(M^{n-1})}{\Huge \wedge }_{\forall
i\in Point^{n-1}\cup \{n-1\}}-DS^{n-1}[t,i]\leq \underline{t_{f}^{_{i+1}}}%
+..+\underline{t_{f}^{n-1}}+\underline{t}\leq DS^{n-1}[i,t]$%
\end{tabular}%
\newline
\right. $\newline
}

{\small {According to the hypotheses of the Clause 2, as we have }$%
\left\rceil DS^{n-1}\right\lceil \subseteq \left\rceil \widetilde{Ds^{n-1}}%
\right\lceil $ then the following properties hold: }

{\small $\left\{ 
\begin{tabular}{l}
$P_{4}:\forall i\in point^{n-1},DS^{n-1}[n-1,i]\leq \widetilde{DS^{n-1}}%
[n-1,i]\quad $ \\ 
$P_{5}:\forall i\in point^{n-1},DS^{n-1}[i,n-1]\leq \widetilde{DS^{n-1}}%
[i,n-1]$ \\ 
$P_{6}:\forall i\in point^{n-1}\cup \{n-1\},\forall t\in Te(M^{n-1}),\quad
DS^{n-1}[t,i]\leq \widetilde{DS^{n-1}}[t,i]$ \\ 
$P_{7}:\forall i\in point^{n-1}\cup \{n-1\},\forall t\in Te(M^{n-1}),\quad
DS^{n-1}[i,t]\leq \widetilde{DS^{n-1}}[i,t]$%
\end{tabular}%
\right. $ }

{\small Let assume now the systems }$\widetilde{{\small DS^{n}}}${\small \ \
and }${\small DS^{n}}${\small \ associated respectively with the classes $%
\widetilde{E_{c}^{n}}$\ and $E^{n}$ obtained after firing the transition }$%
{\small t}_{{\small f}}^{{\small n}}${\small :}

{\small $\left\{ 
\begin{tabular}{l}
$F_{1}:{\Huge \wedge }_{\forall i\in Point^{n}}-\widetilde{DS^{n}}[n,i]\leq 
\underline{t_{f}^{_{i+1}}}+..+\underline{t_{f}^{n}}\leq \widetilde{DS^{n}}%
[i,n]$\newline
\\ 
$F_{2}:{\Huge \wedge }_{\forall i\in Point^{n}\cup \{n\}}${\Huge $\wedge $}$%
_{\forall t^{\prime }\in Te(M^{n})}-\widetilde{DS^{n}}[t^{\prime },i]\leq 
\underline{t_{f}^{_{i+1}}}+..+\underline{t_{f}^{n}}+\underline{t}^{\prime
}\leq \widetilde{DS^{n}}[i,t^{\prime }]$%
\end{tabular}%
\newline
\right. $\newline
}

{\small $\left\{ 
\begin{tabular}{l}
$F_{3}:{\Huge \wedge }_{\forall i\in Point^{n}}-DS^{n}[n-1,i]\leq \underline{%
t_{f}^{_{i+1}}}+..+\underline{t_{f}^{n}}\leq DS^{n}[i,n]$\newline
\\ 
$F_{4}:{\Huge \wedge }_{\forall i\in Point^{n}\cup \{n\}}${\Huge $\wedge $}$%
_{\forall t^{\prime }\in Te(M^{n})}-DS^{n}[t^{\prime },i]\leq \underline{%
t_{f}^{_{i+1}}}+..+\underline{t_{f}^{n}}+\underline{t}^{\prime }\leq
DS^{n}[i,t^{\prime }]$%
\end{tabular}%
\newline
\right. $\newline
}

{\small We need to prove that : }

{\small $\left\{ 
\begin{tabular}{l}
$P_{4}^{\prime }:\forall i\in point^{n},DS^{n}[n,i]\leq \widetilde{DS^{n}}%
[n,i]\quad $ \\ 
$P_{5}^{\prime }:\forall i\in point^{n},DS^{n}[i,n]\leq \widetilde{DS^{n}}%
[i,n]$ \\ 
$P_{6}^{\prime }:\forall i\in point^{n}\cup \{n\},\forall t^{\prime }\in
Te(M^{n}),\quad DS^{n}[t^{\prime },i]\leq \widetilde{DS^{n}}[t^{\prime },i]$
\\ 
$P_{7}^{\prime }:\forall i\in point^{n}\cup \{n\},\forall t^{\prime }\in
Te(M^{n}),\quad DS^{n}[i,t^{\prime }]\leq \widetilde{DS^{n}}[i,t^{\prime }]$%
\end{tabular}%
\right. $}

{\small First of all, we have: $\forall i$ $\in \left[ Ne^{n}\right] -\{n\}$%
, then $i$ $\in \left[ Ne^{n-1}\right] $, namely all persistent transitions
reported at point $(n)$ keep their same enabling point as in the firing
point $(n-1)$. Furthermore,$\forall i\in \left[ Ni^{n}\right] -\{n\}$, then $%
i$ $\in \left[ Ni^{n-1}\right] .$ In other words, all persistent inhibited
transitions reported at point $(n)$ enjoy the same inhibiting point as for
the point $(n-1)$. Finally,$\forall i\in \left[ Na^{n}\right] -\{n\}$, then $%
i$ $\in \left[ Na^{n-1}\right] ,$ namely all persistent activated
transitions reported at point $(n)$ enjoy the same activating point as for
the point $(n-1)$. Therefore, we have: }$\ {\small (P}_{{\small 8}}^{\prime }%
{\small ):}${\small $\forall i$ $\in Point^{n}-\{n\}$, then $i$ $\in
Point^{n-1}.$}

{\small As described in \textit{Definitions 7}, the computation of the
system $DS^{n}$\ is performed by replacing each variable \underline{$t$}
associated with a persistent activated transition $t\in Ta(M^{n-1})-\left\{
t_{f}^{n}\right\} $ by $\underline{t^{^{\prime }}}+\underline{t_{f}^{n}}.$
However, each variable\ $\underline{t}$ connected to a persistent inhibited
transition is replaced with $\underline{t^{^{\prime }}}.$ \ The coefficients
of $DS^{n}$\ are determined by intersection of the inequalities of
predecessor systems in the sequence, namely $D_{a}^{n-1}$ and $DS^{n-1}$.}

\begin{itemize}
\item {\small Let us determine first the proprerties }$P_{4}^{\prime }$%
{\small \ and }$P_{5}^{\prime }.${\small \ To this end, we restrain our
constraint manipulations by summing only the inequalities of $B_{3}:$$%
\forall t\in Ta(M^{n-1})$ $\quad $\underline{$t_{f}^{n}$}$\leq t$ with the
right part of $C_{4}$, we obtain: }\newline
{\small $\forall t\in Ta(M^{n-1})\quad $$\underline{t_{f}^{_{i+1}}}+..+%
\underline{t_{f}^{n-1}}+$\underline{$t_{f}^{n}$}$+\underline{t}\leq
DS^{n-1}[i,t]+\underline{t}.$ }\newline
{\small Let us remove the variable $\underline{t}$\ from both parts of the
previous inequalities:}\newline
{\small $\underline{t_{f}^{_{i+1}}}+..+\underline{t_{f}^{n-1}}+$\underline{$%
t_{f}^{n}$}$\leq \underset{\forall t\in Ta(M^{n-1})}{MIN}DS^{n-1}[i,t]$ }%
\newline
{\small On the other side, let us consider the left part of the constraint $%
C_{4}$\ while assuming $\underline{t}$\ =\underline{$t_{f}^{n}$}, we obtain: 
$-DS^{n-1}[t_{f}^{n},i]\leq \underline{t_{f}^{_{i+1}}}+..+\underline{%
t_{f}^{n}}$ }\newline
{\small Hence, we determine that :\ }\newline
{\small \ $-DS^{n-1}[t_{f}^{n},i]\leq -DS^{n}[n,i]\leq \underline{%
t_{f}^{_{i+1}}}+..+\underline{t_{f}^{n}}\leq DS^{n}[i,n]\leq \underset{%
\forall t\in Ta(M^{n-1})}{MIN}DS^{n-1}[i,t]$. }\newline
{\small Then by using the properties }${\small P}_{{\small 4}}{\small ..P}_{%
{\small 6}}${\small , we deduce :}\newline
{\small $-\widetilde{DS^{n-1}}[t_{f}^{n},i]\leq -DS^{n}[n,i]\leq \underline{%
t_{f}^{_{i+1}}}+..+\underline{t_{f}^{n}}\leq DS^{n}[i,n]\leq $}$\lambda
^{n-1}[i]${\small . }\newline
{\small Then according to Proposition 1, we prove }${\small P}_{{\small 4}%
}^{\prime }${\small \ and }${\small P}_{{\small 5}}^{\prime }${\small : }%
\newline
{\small $-\widetilde{DS^{n}}[n,i]\leq -DS^{n}[n,i]\leq \underline{%
t_{f}^{_{i+1}}}+..+\underline{t_{f}^{n}}\leq DS^{n}[i,n]\leq \widetilde{%
DS^{n}}[i,n]$.}
\end{itemize}

{\small To determine now the proprerties }${\small P}_{{\small 5}}^{\prime }$%
{\small \ and }${\small P}_{{\small 6}}^{\prime }${\small \ we have to
consider first the status of the transition $t^{\prime }$\ at the firing
point $(n)$. }

\begin{itemize}
\item {\small Case where $t^{\prime }$\ is newly enabled for $M^{n}:$\
Therefore the variable $t^{\prime }$\ is new in $DS^{n}$\ and has not been
obtained by renaming another variable of $DS^{n-1}$. So by intersection of
the constraints of }${\small tmin(t}^{\prime }{\small )\leq \underline{t}%
^{\prime }\leq tmax(t}^{\prime }{\small )}${\small \ \ and $F_{3},$\ we
determine:}\newline
{\small $-DS_{n}[n,i]+tmin(t^{\prime })\leq \underline{t_{f}^{_{i+1}}}+..+%
\underline{t_{f}^{n}}+\underline{t}^{\prime }\leq DS^{n}[i,n]+$}${\small %
tmax(t}^{\prime }{\small ).}$\newline
{\small Then, according to Proposition 1 and by using the properties }$%
{\small P}_{{\small 4}}{\small ..P}_{{\small 6}}${\small , we prove }$%
{\small P}_{{\small 5}}^{\prime }${\small \ and }${\small P}_{{\small 6}%
}^{\prime }${\small :$-\widetilde{DS^{n}}[t^{\prime },n]\leq
-DS^{n}[t^{\prime },n]\leq \underline{t_{f}^{_{i+1}}}+..+\underline{t_{f}^{n}%
}+\underline{t}^{\prime }\leq DS^{n}[i,t^{\prime }]\leq \widetilde{DS^{n}}%
[i,t^{\prime }]$. \medskip }

\item {\small Case where $t^{\prime }$\ is persistent for $M^{n}:$\
Therefore, }${\small r=}${\small $Ne^{n}(t)=Ne^{n-1}(t)\neq n$ and the
variable $t^{\prime }$\ has been obtained by renaming another variable of $%
D^{n-1}$. Let us assume }${\small s=}${\small $Ni^{n-1}(t)$ and $%
p=Na^{n-1}(t).$}

{\small \underline{Case 1: $\forall i\in Point^{n}-\left\{ n\right\} $} : We
should consider the status of the original variable $t$\ in $E^{n-1}.$%
\medskip\ }

\begin{itemize}
\item {\small If $t$ is activated for $M^{n-1}$, then the variable 
\underline{$t$}\ was renamed by $\underline{t^{^{\prime }}}$ in $DS^{n}$\
and we have $t=$\ $\underline{t^{^{\prime }}}+\underline{t_{f}^{n}}. $}%
\newline
{\small Let us consider the constraint $F_{3}$\ while assuming the points }$%
{\small i,}${\small \ $p$\ and $s;$ we obtain :}\newline
{\small $(G_{1}):$ $-DS^{n}[n,i]\leq \underline{t_{f}^{_{i+1}}}+..+%
\underline{t_{f}^{n}}\leq DS^{n}[i,n].$ }\newline
{\small $(G_{2}):$ $-DS^{n}[n,p]\leq \underline{t_{f}^{_{p+1}}}+..+%
\underline{t_{f}^{n}}\leq DS^{n}[p,n].$ }\newline
{\small $(G_{3}):$ $-DS^{n}[n,s]\leq \underline{t_{f}^{_{s+1}}}+..+%
\underline{t_{f}^{n}}\leq DS^{n}[s,n].$ }\newline

{\small Now let us consider the systems $DS^{\mathit{s}}$ and $DS^{\mathit{i}%
}$ computed respectively at points }${\small s}${\small \ and }${\small i}$%
{\small , where we deal with the constraint of type }${\small C}_{{\small 4}%
} ${\small : }\newline
{\small If ($i\leq s$) this means that $t$ was inhibited at point $(s)$
after having already reached the point $(i)$. Therefore, we have: $i\in
Point^{s}$ and $-DS^{\mathit{s}}[t,i]\leq \underline{t_{f}^{_{i+1}}}+..+%
\underline{t_{f}^{s}}+\underline{t}%
%TCIMACRO{\U{b0}}%
%BeginExpansion
{{}^\circ}%
%EndExpansion
\leq DS^{\mathit{s}}[i,t];$ where \underline{$t%
%TCIMACRO{\U{b0}}%
%BeginExpansion
{{}^\circ}%
%EndExpansion
$} is the original name of the variable related to transition $t$ in $E^{s}$%
. Hence, as }${\small t}${\small \ is inhibited in the point interval }$%
{\small [s,p]}${\small \ we replace \underline{$t%
%TCIMACRO{\U{b0}}%
%BeginExpansion
{{}^\circ}%
%EndExpansion
$} in the previous constraint with $\underline{t_{f}^{_{p+1}}}+..+\underline{%
t_{f}^{n}}+\underline{t}^{\prime }$ we obtain:}\newline
{\small $(G_{4}):-DS^{\mathit{s}}[t,i]\leq \underline{t_{f}^{_{i+1}}}+..+%
\underline{t_{f}^{s}}+\underline{t_{f}^{_{p+1}}}+..+\underline{t_{f}^{n}}+%
\underline{t}^{\prime }\leq DS^{\mathit{s}}[i,t];$}

{\small Otherwise, if ($i>s$) this means that $t$ was inhibited at point $%
(s) $ before reaching the point $(i)$. Therefore, we have: $i\notin point$}$%
^{{\small s}}${\small \ but $s\in point$}$^{{\small i}}$ {\small and $-DS^{%
\mathit{i}}[t,i]\leq \underline{t}^{\ast }\leq DS^{\mathit{i}}[i,t];$ where 
\underline{$t^{\ast }$} is the original name of the variable related to
transition $t$ in $E^{i}$. Hence, as }${\small t}${\small \ is inhibited in
the point interval }${\small [i,p]}$ {\small we replace }\underline{{\small $%
t$}}$^{\ast }${\small \ in the previous constraint with $\underline{%
t_{f}^{_{p+1}}}+..+\underline{t_{f}^{n}}+\underline{t}^{\prime }$ we obtain:}%
\newline
{\small $(G_{5}):-DS^{\mathit{i}}[t,i]\leq \underline{t_{f}^{_{p+1}}}+..+%
\underline{t_{f}^{n}}+\underline{t}^{\prime }\leq DS^{\mathit{i}}[i,t];$}

{\small Note recalling that all the variables \underline{$t$}$,\underline{%
t^{\prime }},\underline{t%
%TCIMACRO{\U{b0}}%
%BeginExpansion
{{}^\circ}%
%EndExpansion
}$ and \underline{$t^{\ast }$} relate to the same occurence of the
transition }${\small t}${\small \ since }${\small t}${\small \ remains
persistently enabled in the firing sequence till the point $(n)$. }

{\small Case ($i\leq s$): By summing }${\small G}_{{\small 4}}${\small \ and 
}${\small G}_{{\small 3}}${\small \ and then by intersection with }${\small G%
}_{{\small 2}}${\small , we obtain: }

{\small $(H_{1}):-DS^{\mathit{s}}[t,i]-DS^{\mathit{n}}[p,n]-DS^{\mathit{n}%
}[n,s]\leq -DS^{\mathit{n}}[t,i]\leq \underline{t_{f}^{_{i+1}}}+..+%
\underline{t_{f}^{n}}+\underline{t}^{\prime }\leq DS^{\mathit{n}}[i,t]\leq
DS^{\mathit{s}}[i,t]+DS^{\mathit{n}}[n,p]+DS^{\mathit{n}}[s,n];$}

{\small Case ($s<i$): By summing }${\small G}_{{\small 5}}${\small \ and }$%
{\small G}_{{\small 1}}${\small \ and then by intersection with }${\small G}%
_{{\small 2}}${\small , we obtain: }

{\small $(H_{2}):-DS^{\mathit{i}}[t,i]-DS^{\mathit{n}}[p,n]-DS^{\mathit{n}%
}[n,i]\leq -DS^{\mathit{n}}[t,i]\leq \underline{t_{f}^{_{i+1}}}+..+%
\underline{t_{f}^{n}}+\underline{t}^{\prime }\leq DS^{\mathit{n}}[i,t]\leq
DS^{\mathit{i}}[i,t]+DS^{\mathit{n}}[n,p]+DS^{\mathit{n}}[i,n];$}\newline
{\small On the other hand, let us consider now the constraint $C_{4},$\ we
have:\newline
$-DS^{n-1}[t,i]\leq \underline{t_{f}^{_{i+1}}}+..+\underline{t_{f}^{n-1}}+%
\underline{t}\leq DS^{n-1}[i,t]$}\newline
{\small We put $\underline{t}$ $=$\ $\underline{t^{^{\prime }}}+\underline{%
t_{f}^{n}},$ we obtain $:$ $-DS^{n-1}[t,i]\leq \underline{t_{f}^{_{i+1}}}+..+%
\underline{t_{f}^{n}}+\underline{t}^{\prime }\leq DS^{n-1}[i,t]$}\newline
{\small $(H_{3}):$ $-DS^{n-1}[t,i]\leq -DS^{n}[t,i]\leq \underline{%
t_{f}^{_{i+1}}}+..+\underline{t_{f}^{n}}+\underline{t}^{\prime }\leq
DS^{n}[i,t]\leq DS^{n-1}[i,t]$}\newline
{\small In other respects, by intersection of }${\small B}_{{\small 3}}$%
{\small \ and }${\small C}_{{\small 4}}${\small , and then summing with }$%
{\small G}_{{\small 1}}${\small , we obtain:}\newline
{\small $(H_{4}):-DS^{\mathit{n}}[n,i]-MIN(0,DS^{\mathit{n-1}%
}[t,n-1]+\lambda ^{\mathit{n-1}}[n-1])\leq -DS^{\mathit{n}}[t,i]\leq 
\underline{t_{f}^{_{i+1}}}+..+\underline{t_{f}^{n}}+\underline{t}^{\prime
}\leq DS^{\mathit{n}}[i,t]\leq DS^{\mathit{n}}[i,n]+DS^{\mathit{n-1}%
}[n-1,t]+DS^{\mathit{n-1}}[t_{f}^{n},n-1];$}\newline
{\small Finally, from }${\small H}_{{\small 1}}{\small ,H}_{{\small 2}}%
{\small ,H}_{{\small 3}}${\small \ and }${\small H}_{{\small 4}}${\small ,
by using previous established properties\ and according to proposition 1, we
determine }${\small P}_{{\small 6}}^{\prime }${\small \ and }${\small P}_{%
{\small 7}}^{\prime }${\small \ :}\newline
{\small $-DS^{\mathit{n}}[t,i]\leq -\widetilde{DS^{\mathit{n}}}[t,i]\leq 
\underline{t_{f}^{_{i+1}}}+..+\underline{t_{f}^{n}}+\underline{t}^{\prime
}\leq DS^{\mathit{n}}[i,t]\leq \widetilde{DS^{\mathit{n}}}[i,t].\medskip $}

\item {\small If $t$ is inhibited for $M^{n-1}$, then $t$\ was renamed by $%
\underline{t^{^{\prime }}}$ in $DS^{n}$\ and we have $t=$\ $\underline{%
t^{^{\prime }}}.$}\newline
{\small Let us consider first the constraint $C_{4},$\ we have:\newline
$-DS^{n-1}[t,i]\leq \underline{t_{f}^{_{i+1}}}+..+\underline{t_{f}^{n-1}}+%
\underline{t}\leq DS^{n-1}[i,t]$}\newline
{\small We put $\underline{t}$ $=$\ $\underline{t^{^{\prime }}},$ we obtain }%
${\small (G}_{{\small 1}}^{\prime }{\small ):}${\small \ $-DS^{n-1}[t,i]\leq 
\underline{t_{f}^{_{i+1}}}+..+\underline{t_{f}^{n-1}}+\underline{t}^{\prime
}\leq DS^{n-1}[i,t].$}\newline
{\small In other respects, by intersection of }${\small B}_{{\small 3}}$%
{\small \ with }${\small C}_{{\small 4}}${\small , and then summing with }$%
{\small G}_{{\small 1}}^{\prime }${\small , we obtain:}\newline
{\small $(H_{1}^{\prime }):-DS^{\mathit{n-1}}[t,i]-DS^{\mathit{n-1}%
}[t_{f}^{n},n-1]\leq -DS^{\mathit{n}}[t,i]\leq \underline{t_{f}^{_{i+1}}}+..+%
\underline{t_{f}^{n}}+\underline{t}^{\prime }\leq DS^{\mathit{n}}[i,t]\leq
\lambda ^{\mathit{n-1}}[n-1]+DS^{\mathit{n-1}}[i,t];$}

{\small Let us consider the constraint $F_{3}$\ while assuming the points $s$
and }${\small i}${\small , we obtain :}\newline
{\small $(G_{2}^{\prime }):$ $-DS^{n}[n,s]\leq \underline{t_{f}^{_{s+1}}}+..+%
\underline{t_{f}^{n}}\leq DS^{n}[s,n].$ }\newline
{\small $(G_{3}^{\prime }):$ $-DS^{n}[n,i]\leq \underline{t_{f}^{_{i+1}}}+..+%
\underline{t_{f}^{n}}\leq DS^{n}[i,n].$ }\newline

{\small Now let us consider the systems $DS^{\mathit{s}}$ and $DS^{\mathit{i}%
}$ computed respectively at point }${\small s}${\small \ and }${\small i}$%
{\small , where we deal with the constraint of type }${\small C}_{{\small 4}%
} ${\small : }\newline
{\small If ($i\leq s$) this means that $t$ was inhibited at point $(s)$
after having already reached the point $(i)$ and still remains persistently
inhibited till point }${\small (n)}${\small . Therefore, we have: $i\in
Point^{s}$ and $-DS^{\mathit{s}}[t,i]\leq \underline{t_{f}^{_{i+1}}}+..+%
\underline{t_{f}^{s}}+\underline{t}%
%TCIMACRO{\U{b0}}%
%BeginExpansion
{{}^\circ}%
%EndExpansion
\leq DS^{\mathit{s}}[i,t];$ where \underline{$t%
%TCIMACRO{\U{b0}}%
%BeginExpansion
{{}^\circ}%
%EndExpansion
$} is the original name of the variable related to transition $t$ in $E^{s}$%
. Hence, as }${\small t}${\small \ is inhibited in the point interval }$%
{\small [s,n]}${\small \ we replace \underline{$t%
%TCIMACRO{\U{b0}}%
%BeginExpansion
{{}^\circ}%
%EndExpansion
$} in the previous constraint with $\underline{t}^{\prime }$ we obtain:}%
\newline
{\small $(G_{4}^{\prime }):-DS^{\mathit{s}}[t,i]\leq \underline{%
t_{f}^{_{i+1}}}+..+\underline{t_{f}^{s}}++\underline{t}^{\prime }\leq DS^{%
\mathit{s}}[i,t];$}

{\small Otherwise, if ($i>s$) this means that $t$ was inhibited at point $%
(s) $ before reaching the point $(i)$ and still remains persistently
inhibited till point }${\small n}$. {\small Therefore, we have: $i\notin
point$}$^{{\small s}}${\small \ but $s\in point$}$^{{\small i}}$ {\small and 
$-DS^{\mathit{i}}[t,i]\leq \underline{t}^{\ast }\leq DS^{\mathit{i}}[i,t];$
where \underline{$t^{\ast }$} is the original name of the variable related
to transition $t$ in $E^{i}$. Hence, as }${\small t}${\small \ is inhibited
in the point interval }${\small [i,n]}$ {\small we replace }\underline{%
{\small $t$}}$^{\ast }${\small \ in the previous constraint with $\underline{%
t}^{\prime }$ we obtain: $(G_{5}^{\prime }):-DS^{\mathit{i}}[t,i]\leq 
\underline{t}^{\prime }\leq DS^{\mathit{i}}[i,t];$}

{\small Case ($i\leq s$): By summing }${\small G}_{{\small 4}}^{\prime }$%
{\small \ and }${\small G}_{{\small 2}}^{\prime }${\small \ we obtain: }

{\small $(H_{2}^{\prime }):-DS^{\mathit{s}}[t,i]-DS^{\mathit{n}}[n,s]\leq
-DS^{\mathit{n}}[t,i]\leq \underline{t_{f}^{_{i+1}}}+..+\underline{t_{f}^{n}}%
+\underline{t}^{\prime }\leq DS^{\mathit{n}}[i,t]\leq DS^{\mathit{s}%
}[i,t]+DS^{\mathit{n}}[s,n];$}

{\small Case ($s<i$): By summing }${\small G}_{{\small 5}}^{\prime }${\small %
\ and }${\small G}_{{\small 3}}^{\prime }${\small , we obtain: }

{\small $(H_{3}^{\prime }):-DS^{\mathit{i}}[t,i]-DS^{\mathit{n}}[n,i]\leq
-DS^{\mathit{n}}[t,i]\leq \underline{t_{f}^{_{i+1}}}+..+\underline{t_{f}^{n}}%
+\underline{t}^{\prime }\leq DS^{\mathit{n}}[i,t]\leq DS^{\mathit{i}%
}[i,t]+DS^{\mathit{n}}[i,n];$}\newline
{\small Finally, from }${\small H}_{{\small 1}}^{\prime }{\small ,H}_{%
{\small 2}}^{\prime }{\small ,}${\small \ and }${\small H}_{{\small 3}%
}^{\prime }${\small , by using previous established properties and according
to proposition 1, we determine the properties :}\newline
{\small $-DS^{\mathit{n}}[t,i]\leq -\widetilde{DS^{\mathit{n}}}[t,i]\leq 
\underline{t_{f}^{_{i+1}}}+..+\underline{t_{f}^{n}}+\underline{t}^{\prime
}\leq DS^{\mathit{n}}[i,t]\leq \widetilde{DS^{\mathit{n}}}[i,t].\medskip $}
\end{itemize}
\end{itemize}

{\small \underline{Case $i=n$:\newline
} \ First of all, it should be noticed that we have by definition:}\newline
\ \ \ \ \ \ \ \ \ \ \ \ \ \ {\small $\widetilde{DS^{n}}[n,t]=\widetilde{%
D_{c}^{n}}[\bullet ,t]$ and $\widetilde{DS^{n}}[t,n]=\widetilde{D_{c}^{n}[}%
t,\bullet ].$ Moreover, we have :\ $DS^{n}[n,t]=\overrightarrow{D^{n}}%
[\bullet ,t]$ and $DS^{n}[t,n]=\overrightarrow{D^{n}}[t,\bullet ].$ }

\begin{itemize}
\item 
\begin{itemize}
\item {\small If $t$ is activated for $M^{n-1}$: \ From we properties }$%
{\small P}_{{\small 1}}{\small ..P}_{{\small 4}}${\small \ we have : $\beta
^{n-1}[t]\leq \widetilde{\beta _{c}^{n-1}}[t]\leq \widetilde{\beta ^{n-1}}%
[t].$}\newline
{\small As already established in \cite{abdelli} and shown in Definition.5,
when manipulating exclusively the DBM constraints of the normalized systems }%
$\overrightarrow{{\small D^{n-1}}}${\small \ or }$\widetilde{{\small %
D_{c}^{n-1}}}${\small \ or }$\widetilde{{\small D^{n-1}}}${\small \ we
obtain: }\newline
{\small $(L_{1}):-\widetilde{\beta ^{n-1}}[t]$\ $\leq \underline{t}^{\prime
}\leq $}$\widetilde{{\small D^{n-1}}}${\small $[t_{f}^{n},t].\ $}\newline
{\small $(L_{2}):-$}${\small \beta ^{n-1}{}}${\small $[t]$\ $\leq \underline{%
t}^{\prime }\leq $}$\overrightarrow{{\small D^{n-1}}}${\small $%
[t_{f}^{n},t].\ $}\newline
{\small $(L_{3}):-$}$\widetilde{{\small \beta _{c}^{n-1}}}${\small ${}[t]$\ $%
\leq \underline{t}^{\prime }\leq \widetilde{D_{c}^{n-1}}[t_{f}^{n},t].\ $}%
\newline
{\small Let us consider now the constraints of $F_{3}$\ and $F_{4}$ with $%
i=r $ : }\newline
{\small $-DS^{n}[t,r]\leq \underline{t_{f}^{_{r+1}}}+..+\underline{t_{f}^{n}}%
+\underline{t}^{\prime }\leq DS^{n}[r,t]$}\newline
{\small $-DS^{n}[n,r]\leq \underline{t_{f}^{_{r+1}}}+..+\underline{t_{f}^{n}}%
\leq DS^{n}[r,n]$}\newline
{\small By intersection of the previous constraints we obtain:}\newline
{\small $(L_{4}):-MIN(0,DS^{n}[t,r]+DS^{n}[r,n]\leq \underline{t}^{\prime
}\leq DS^{n}[r,t]+DS^{n}[n,r]$}\newline
{\small From }${\small L}_{{\small 1}}{\small ,..L}_{{\small 4}}${\small ,
Proposition 1 and D\'{e}finition 5. Then by using already established
properties, we deduce : }\newline
${\small -}${\small $\widetilde{D^{n}}[t,\bullet ]\leq $ }${\small -}$%
{\small $\widetilde{DS^{n}}[t,n]\leq -DS^{n}[t,n]\leq \underline{t}^{\prime
}\leq DS^{n}[n,t]\leq $ $\widetilde{DS^{n}}[n,t]\leq $ $\widetilde{D^{n}}%
[\bullet ,t]$}.\medskip

\item {\small If $t$ is inhibited for $M^{n-1}.$} {\small As already
established in \cite{abdelli} and shown in Definition.5, \ when dealing
exclusively with the DBM constraints of the normalized systems }$%
\overrightarrow{{\small D^{n-1}}}${\small \ or }$\widetilde{{\small %
D_{c}^{n-1}}}${\small \ or }$\widetilde{{\small D^{n-1}}}${\small \ we
obtain: }\newline
{\small $(L_{1}^{\prime }):-MIN$}$\left( {\small \widetilde{D^{n-1}}%
[t,\bullet ],}\text{ }{\small \widetilde{D^{n-1}}[t_{f}^{n},\bullet ]+\ 
\widetilde{\beta ^{n-1}}[t]}\right) ${\small $\leq \underline{t}^{\prime
}\leq MIN$}$\left( \widetilde{{\small D^{n-1}}}{\small [\bullet ,t],\ }%
\widetilde{{\small D^{n-1}}}{\small [t_{f}^{n},t]+\ }\widetilde{{\small %
\beta ^{n-1}}}{\small [\bullet ]}\right) ${\small $.\ $}\newline
{\small $(L_{2}^{\prime }):-MIN$}$\left( {\small \widetilde{DS^{n-1}}[t,n-1],%
}\text{ }{\small \widetilde{DS^{n-1}}[t_{f}^{n},n-1]+\ \widetilde{\beta
_{c}^{n-1}}[t]}\right) ${\small $\leq \underline{t}^{\prime }\leq $ $\ $}%
\newline
\ \ \ \ \ \ \ \ \ \ \ \ \ \ \ \ \ \ \ \ \ \ \ \ \ \ {\small $MIN$}$\left( 
\widetilde{{\small DS}^{{\small n-1}}}{\small [n-1,t],\ }\widetilde{{\small D%
}_{{\small c}}^{{\small n-1}}}{\small [t_{f}^{n},t]+\ \lambda
^{n-1}[n-1]\medskip }\right) .$\newline
{\small $(L_{3}^{\prime }):\ -MIN$}$\left( {\small DS^{n-1}[t,n-1],}\text{ }%
{\small DS^{n-1}[t_{f}^{n},n-1]+\ \beta ^{n-1}[t]}\right) ${\small $\leq 
\underline{t}^{\prime }\leq $}\newline
{\small $MIN$}$\left( {\small DS^{n-1}[n-1,t],\ DS^{n-1}[t_{f}^{n},t]+\
\beta ^{n-1}[\bullet ]}\right) ${\small $.\ $}\newline
{\small Notice that we have : $\widetilde{\beta _{c}^{n-1}}[\bullet
]=\lambda ^{n-1}[n-1]$\medskip .}\newline
{\small Let us consider now the constraints of $F_{3}$\ and $F_{4}$ with $%
i=r $ : }\newline
{\small $-DS^{n}[t,r]\leq \underline{t_{f}^{_{r+1}}}+..+\underline{t_{f}^{n}}%
+\underline{t}^{\prime }\leq DS^{n}[r,t]$}\newline
{\small \ $-DS^{n}[n,r]\leq \underline{t_{f}^{_{r+1}}}+..+\underline{%
t_{f}^{n}}\leq DS^{n}[r,n]$}\newline
{\small By intersection of the previous constraints we obtain:}\newline
{\small $(L_{4}^{\prime }):-MIN(0,DS^{n}[t,r]+DS^{n}[r,n]\leq \underline{t}%
^{\prime }\leq DS^{n}[r,t]+DS^{n}[n,r]$}\newline
{\small From }${\small L}_{{\small 1}}^{\prime }{\small ,..L}_{{\small 4}%
}^{\prime }${\small , Proposition 1 and D\'{e}finition 5. Then by using
already established properties, we deduce : }\newline
${\small -}${\small $\widetilde{D^{n}}[t,\bullet ]\leq $ }${\small -}$%
{\small $\widetilde{DS^{n}}[t,n]\leq -DS^{n}[t,n]\leq \underline{t}^{\prime
}\leq DS^{n}[n,t]\leq $ $\widetilde{DS^{n}}[n,t]\leq $ $\widetilde{D^{n}}%
[\bullet ,t]$}\newline
\newline
\end{itemize}
\end{itemize}

{\small We prove the properties }${\small P}_{{\small 4}}^{\prime }{\small %
,..P}_{{\small 7}}^{\prime }${\small , therefore the system }${\small 
\widetilde{DS_{n}}}${\small \ \ is an over-approximation of the system }$%
{\small DS_{n}}${\small $.$\ }

{\small We need to establish now that :}

{\small $\left\{ 
\begin{tabular}{l}
$P_{1}^{\prime }:\forall t_{_{1}}\neq t_{_{2}}\in Te(M^{n}),\quad 
\overrightarrow{D^{n}}[t_{_{1}},t_{_{2}}]\leq \widetilde{D_{c}^{n}}%
[t_{_{1}},t_{_{2}}]\leq \widetilde{D^{n}}[t_{_{1}},t_{_{2}}].\quad $ \\ 
$P_{2}^{\prime }:\forall t\in Te(M^{n}),\quad \overrightarrow{D^{n}}%
[t,\bullet ]\leq \widetilde{D_{c}^{n}}[t,\bullet ]\leq \widetilde{D^{n}}%
[t,\bullet ].\quad $ \\ 
$P_{3}^{\prime }:\forall t\in Te(M^{n}),\quad \overrightarrow{D^{n}}[\bullet
,t]\leq \widetilde{D_{c}^{n}}[\bullet ,t]\leq \widetilde{D^{n}}[\bullet
,t].\quad $%
\end{tabular}%
\right. $ }

\begin{itemize}
\item {\small As we have: $\widetilde{DS^{n}}[n,t]=\widetilde{D_{c}^{n}}%
[\bullet ,t]$ ; $\widetilde{DS^{n}}[t,n]=\widetilde{D_{c}^{n}[}t,\bullet ]$
and\ $DS^{n}[n,t]=\overrightarrow{D^{n}}[\bullet ,t]$ and $DS^{n}[t,n]=%
\overrightarrow{D^{n}}[t,\bullet ],$ we deduce easily from previous results
properties }${\small P}_{{\small 2}}^{\prime }${\small \ and }${\small P}_{%
{\small 3}}^{\prime }${\small .}

\item {\small Let us prove the property }${\small P}_{{\small 1}}^{\prime }$%
{\small . \ For this effect, we have shown in \cite{Abdelli un} that the
algorithm of Definition 5 allows to compute an overapproximation of the
system}$\ ${\small $D^{n}:$ $\forall t_{_{1}}\neq t_{_{2}}\in
Te(M^{n}),\quad \overrightarrow{D^{n}}[t_{_{1}},t_{_{2}}]\leq \widetilde{%
D^{n}}[t_{_{1}},t_{_{2}}].\quad $On the other side, we notice from
Definition 8 that the system $\widetilde{D_{c}^{n}}\ $is computed by using
much precise formulae then those used in Definition 5 to compute the system $%
\widetilde{D^{n}}$. In actual fact, each coefficient of the system $%
\widetilde{D_{c}^{n}}$ is determined as a minimum of two values. The first
one is obtained by maniplulating the constraints of $\widetilde{D_{c}^{n-1}}$
and uses the same formulae as for computing the systems $\widetilde{D^{n}}.$
The second value is obtained by manipulating the coefficients of the system $%
\widetilde{DS^{n}}.$ Therfore, we have $\forall t_{_{1}}\neq t_{_{2}}\in
Te(M^{n}),\quad \widetilde{D_{c}^{n}}[t_{_{1}},t_{_{2}}]\leq \widetilde{D^{n}%
}[t_{_{1}},t_{_{2}}].$}\newline

{\small It is noteworthy that in the context of the exact graph, the system }%
${\small DS}^{n}$ {\small is redundent relatively to the system }${\small D}%
^{n}${\small \ }$;${\small \ the latter does not restric the firing space of 
}${\small D}^{n}${\small . Assuming that, let us consider the constraints }$%
{\small F}_{{\small 4}}${\small \ involving the two enabled transitions }$%
{\small t}_{{\small 1}}${\small \ and }${\small t}_{2}${\small \ for all
points pertaining to }${\small Point}^{{\small n}}\cup \{n\}.${\small \ }%
\newline
$(K_{1}):${\small $\underset{i\in Point^{n}\cup \{n\}}{\wedge }$ $%
-DS^{n}[t_{1},i]\leq \underline{t_{f}^{_{i+1}}}+..+\underline{t_{f}^{n}}%
+t_{1}\leq DS^{n}[i,t_{1}]$}\newline
$(K_{2}):${\small $\underset{i\in Point^{n}\cup \{n\}}{\wedge }$ $%
-DS^{n}[t_{2},i]\leq \underline{t_{f}^{_{i+1}}}+..+\underline{t_{f}^{n}}%
+t_{2}\leq DS^{n}[i,t_{2}]$}\newline
{\small By intersection between the constraints }${\small K}_{{\small 1}}$%
{\small \ and those of }${\small K}_{{\small 2}}${\small \ we obtain :}%
\newline
$(M_{1}):-${\small $\underset{i\in Point^{n}\cup \{n\}}{MIN}$ $%
(DS^{n}[i,t_{1}]$}$+{\small DS^{n}[t_{2},i]})${\small $\leq t_{2}-t_{1}\leq 
\underset{i\in Point^{n}\cup \{n\}}{MIN}(DS^{n}[i,t_{2}]$}$+${\small $%
DS^{n}[t_{1},i])$}\newline
{\small Notice that we have $\alpha ^{n}[t,t^{\prime }]$\ =$\underset{i\in
Point^{n}\cup \{n\}}{MIN}\left( \widetilde{DS^{n}}[i,t^{\prime }]+\widetilde{%
DS^{n}}[t,i]\right) .$}

{\small From previous established properties, Propodition 1, we determine }$%
{\small P}_{{\small 1}}^{\prime }.${\small Consequently, we prove that $%
\left\rceil D^{n}\right\lceil \subseteq \left\rceil \widetilde{D_{c}^{n}}%
\right\lceil $}${\small \subseteq \left\rceil \widetilde{D^{n}}\right\lceil .%
}$
\end{itemize}

\end{document}